\documentclass{article}
\usepackage[utf8]{inputenc}
\usepackage[margin=1in]{geometry}
\usepackage[english]{babel}
\usepackage{amsmath}
\usepackage{amssymb}
\usepackage{mathtools}
\usepackage{subcaption}
\usepackage{authblk}
\usepackage{float}
\usepackage[ruled,vlined]{algorithm2e}
\usepackage[backend=biber,style=nature,citestyle=nature]{biblatex}
\usepackage[symbol]{footmisc}

\DeclareMathOperator{\inv}{inv}
\DeclarePairedDelimiterX\set[1]\lbrace\rbrace{#1}
\DeclarePairedDelimiter\abs{\lvert}{\rvert}

\DeclareUnicodeCharacter{03C1}{$\rho$}
\DeclareUnicodeCharacter{2080}{\textsubscript{0}}

\title{Evaluation of statistical approaches for association testing in noisy drug screening data}

\author[1,2* ]{Petr Smirnov}
\author[1,2* ]{Ian Smith}
\author[2]{Zhaleh Safikhani}
\author[2]{Wail Ba-alawi}
\author[2]{Farnoosh Khodakarami}
\author[3]{Eva Lin}
\author[3]{Yihong Yu}
\author[3]{Scott Martin}
\author[4]{Janosch Ortmann}
\author[5,6,7,8]{Tero Aittokallio}
\author[9]{Marc Hafner}
\author[1,2,10]{Benjamin Haibe-Kains}
\affil[1]{Department of Medical Biophysics, University of Toronto, Toronto, Canada}
\affil[2]{Princess Margaret Cancer Center, University Health Network, Toronto, Canada}
\affil[3]{Department of Discovery Oncology, Genentech Inc. South San Francisco, USA}
\affil[4]{Département d’analytique, opérations et technologies de l’information, École des sciences de la gestion, Université du Québec à Montréal, Montréal, Canada}
\affil[5]{iCAN Digital Precision Cancer Medicine Flagship, Helsinki, Finland}
\affil[6]{Institute for Molecular Medicine Finland (FIMM), Helsinki Institute of Life Science (HiLIFE), University of Helsinki, Helsinki, Finland}
\affil[7]{Department of Computer Science, Helsinki Institute for Information Technology (HIIT), Aalto University, Espoo, Finland}
\affil[8]{Department of Mathematics and Statistics, University of Turku, Turku, Finland}
\affil[9]{Department of Oncology Bioinformatics, Genentech Inc. South San Francisco, USA}
\affil[10]{Vector Institute, Toronto, Canada}
\affil[*]{These authors contributed equally to this work.}
\date{April 2021}

\begin{document}

\maketitle

\section*{Abstract}
Identifying associations among biological variables is a major challenge in modern quantitative biological research, particularly given the systemic and statistical noise endemic to biological systems. Drug sensitivity data has proven to be a particularly challenging field for identifying associations to inform patient treatment. To address this, we introduce two semi-parametric variations on the commonly used concordance index: the robust concordance index and the kernelized concordance index (rCI, kCI), which incorporate measurements about the noise distribution from the data. We demonstrate that common statistical tests applied to the concordance index and its variations fail to control for false positives, and introduce efficient implementations to compute p-values using adaptive permutation testing. We then evaluate the statistical power of these coefficients under simulation and compare with Pearson and Spearman correlation coefficients. Finally, we evaluate the various statistics in matching drugs across pharmacogenomic datasets. We observe that the rCI and kCI are better powered than the concordance index in simulation and show some improvement on real data. Surprisingly, we observe that the Pearson correlation was the most robust to measurement noise among the different metrics. \\
\newpage

\section*{Introduction}
Modern biological research is often based on the collection and analysis of massive and high-dimensional datasets. Advances in robotics, microfluidics, sequencing, and information technologies have enabled the profiling of both molecular cell states and phenotypes across millions of conditions, with 100s of millions or more associations that can be explored among the measured variables. With so many relationships to consider, filtering out those that are not significant, both in strength and probability of occurring through random chance, is a primary task in analyzing these data. Towards this end, statistical inference and machine learning have become standard tools of a scientist trying to make sense of massive biological data \cite{greeneBigDataBioinformatics2014,chingOpportunitiesObstaclesDeep2018,mooreBioinformatics2007}. \\

Major challenges in the statistical analyses of biological data include the non-standard distributions of observed values and the noisiness of the measurements, especially for high throughput phenotypic measurements. The direct measurements often need to be transformed and/or normalized before the values can be interpreted and compared with results from other experiments, even within the same study. These transformations result in values that are often relative, bounded, highly skewed and/or distributed with heavy tails. Furthermore, the complexity of the experimental design and assays used, as well as the necessary compromises to enable high throughput collection of data (for example, working with small cell numbers and liquid volumes in plates with thousands of wells), make measurement noise unavoidable within these data. Fortunately, by taking replicate measurements of the same conditions, it is possible to directly measure the degree of noise introduced by a given experimental protocol. \\   

A common approach in identifying associations from biological data is computing similarities between vectors of biological measurements, often using measures of association. The most widely used association measure is Pearson's Product Moment Correlation, which assesses the degree of linear association between variables. The standard statistical test of significance for Pearson's correlation makes parametric assumptions, including normality and homoscedasticity of the inputs, and testing may be sensitive to outliers. Non-parametric correlation measures, such as the Spearman rank correlation and its associated tests, can detect nonlinear, monotonic associations but are commonly accepted as less powered to detect significant linear associations. The concordance index (CI), a linear transformation of Kendall's Tau, is a non-parametric association measure with the advantage of being highly adaptable to missing and censored data, where the exact value of a data point is bounded but not precisely known. These non-parametric statistics also come with commonly associated tests for significance, including exact tests (often requiring no ties in the data), and asymptotically correct tests. The asymptotic tests make weak assumptions on the distribution of the data, however, they converge only in the limit of large sample sizes, and rates of convergence are difficult to characterize in practice.  \\ 

In our work, we examine the behaviour of the common correlation coefficients when applied to regimes motivated by high-throughput screens of cancer cell lines measuring viability after treatment with compounds. We show that for realistic sample sizes, common statistical tests applied to the concordance index fail to control for false positives, reinforcing the need for testing against a non-parametric, permutation-derived null distribution. We carefully examine the power of permutation tests of standard correlation measures in simulated settings, including the addition of simulated noise of realistic magnitudes. We also introduce two novel modifications of the concordance index, the robust CI (rCI) and kernelized CI (kCI), which take into account the exact quantification of noise possible through comparing replicates into the assessment of similarity between two vectors of measurements. We characterize these novel statistics through comparison with existing correlation coefficients, and provide efficient algorithms and implementations to compute them. We found that the standard CI is more powerful under a permutation test in detecting non-zero correlations for bounded and skewed distributions. We also surprisingly observed that the Pearson correlation was more robust to measurement noise than the other correlation coefficients investigated, including the proposed rCI and kCI statistics.  \\

\section*{Related Work}

Assessing how correlation coefficients behave when applied to non-normal data and in the presence of noise has been previously investigated. Published studies have examined the power of correlation based permutation tests, as well as compared permutation tests to those based on asymptotic limiting behaviours of the coefficients \cite{chokPearsonSpearmanKendall2010,bisharaTestingSignificanceCorrelation2012,puthEffectiveUsePearson2014}. Work by Bishara and Hitter \cite{bisharaTestingSignificanceCorrelation2012} (with followup work by Puth et al. \cite{puthEffectiveUsePearson2014}) not only shows a comprehensive comparison of Type 1 error control and power between several parametric and resampling-based tests for the Pearson and Spearman correlations, but also contains a review of early simulation work on permutation testing of the Pearson correlation. These previous studies tested a wide range of unbounded distributions, but did not investigate simultaneously skewed and bounded distributions, a situation which arises frequently in analysis of high throughput phenotypic screens. Previous studies (excepting Chok's work \cite{chokPearsonSpearmanKendall2010}) have also left out the Concordance Index/Kendall's Tau correlation or any other pairwise ranking based methods from their comparisons. Previous studies have also investigated the effect of assay/measurement noise \cite{saccentiCorruptionPearsonCorrelation2020} or non-normal distributed data \cite{bisharaReducingBiasError2015} on the bias of the sample Pearson correlation coefficient, but not on the statistical power to detect a significant effect. The robustness of other correlation coefficients to the presence of such measurement noise likewise remains unstudied. Finally, modifications of CI have been proposed previously in literature, both in the context of biological data \cite{Costello2014}, as well as in the computer science literature for assessing information retrieval system performance \cite{cormackPowerBiasSubset2007}. However, these modified CI statistics were used primarily as a performance metric for concordance between predictors and observations, and not as a measure of correlation between two sets of measurements. \\


Our study differs from existing studies in several aspects. First, we investigate larger sample sizes and lower p-value thresholds for significance ($\alpha$ levels), more reflective of the high-throughput assays and the necessary corrections for the multitude of correlations assessed when analyzing modern biological data. Unlike previous studies, our intent is not an exhaustive characterization of different tests for significance of correlation coefficients across a variety of distributions. Rather, we advocate permutation testing as a general method applicable across correlation coefficients, and ask which coefficient is most powerful at detecting a significant (monotonic) effect. We investigate the statistical power only in the normally distributed case and a single bounded and skewed distribution, focusing instead on comparing different correlations and assessing their performance under noise. Our robust and kernelized Concordance Index measures are also unique in using replicate measurements to directly address the effect of noise in the analyzed data, and we evaluate their performance in detecting significant correlations. 

\section*{Motivating Data: Preclinical Drug Sensitivity Screening}

While the questions asked in this paper are general and not necessarily tied to the analysis of any particular type of biological assay or experiment, our analyses were motivated by applications to preclinical drug sensitivity screening. These screens usually test the ability of compounds to inhibit cell growth and/or induce cell death across a panel of cancer cell line models grown in vitro. Each cell line is treated at several concentrations of compound, and the cell line growth at a particular time-point is compared to a matched untreated control to derive a \% viability value for each concentration level. \\

In the analysis of these data, it is common to fit a Hill Curve to the multi-dose viability measurements of a cancer cell line in response to treatment by a particular drug, as shown in Supplemental Figure \ref{fig:ddrcMetrics}\cite{Prinz2010HillMechanisms, Beam2011OptimizationComputation}. From these dose-response curves, summary metrics are derived that try to capture the sensitivity of a particular cell line to treatment with the compound in a single number. The most commonly used summary metrics are the IC50, the concentration at which the curve crosses 50\%, and the Area Above or Under the Curve (AAC or AUC). The IC50 is a measure of compound potency, while the AAC/AUC averages the signal between potency and maximum efficacy of a compound. The advantage of the IC50 is that it is in micromolar units and can be compared directly between experiments, however, it is not guaranteed to exist in all cases. The AUC/AAC metrics can always be calculated, but they are dependent on the exact concentration ranges chosen for a particular experiment. In practice, many researchers prefer the IC50 measure because of its natural interpretation as a concentration. For examples within our study, we will be using the AAC, as it has been shown to be more consistent between published datasets and does not suffer from missing or truncated values \cite{Haibe-Kains2013InconsistencyStudies, Haverty2016ReproduciblePanels}. As is standard practice \cite{safikhaniChapterPharmacologicalGenetic2016}, we calculate the area above the curve in log concentration space, and normalize the AAC by the total possible area for each experiment determined by the range of concentration measured, leading to values lying in the bounded range between 0 and 1. Note that this is similar to a calculation of mean viability over the concentrations range as a fraction. \\

\section*{Coefficients Considered}
The general problem is to identify monotonic associations between pairs of variables drawn from a range of potential distributions, and to that end, we considered several parametric and non-parametric association statistics. The Concordance Index (CI) is defined as the fraction of pairs of observations that are ordered the same way by two variables, and it is a linear transformation of Kendall's Tau. Formally, for two variables $x$ and $y$, CI is defined as:
\begin{equation}
    CI = \frac{\text{\# Concordant Pairs}}{\text{\# Pairs}} = \frac{2}{N(N-1)}\sum_{i,j}I(x_{i} > x_{j}, y_{i} > y_{j})
\end{equation}
where \textit{N} the number of observations of $x$ and $y$. CI has a range of $[0,1]$, is 0 when the two variables are perfectly anti-correlated, 1 if they are perfectly correlated, and 0.5 in expectation over the space of all possible orderings, i.e., if there is no association between the variables. 

Even though the Concordance Index is a useful non-parametric statistic, we hypothesize that incorporating parametric information about two observations are can increase the robustness of the Concordance Index to noise. For instance, in physical and biological measurements, measurement error often has a characteristic scale smaller than the range of possible measurements. If measurement error is localized, increasing magnitude of the difference between the value of two observations increases the confidence that the two observations are ordered correctly by the measurements. 

The Robust Concordance Index (rCI) is a modification of CI that only considers pairs of points that are sufficiently dissimilar in both variables. We define two thresholds $\delta_{x} \geq 0$ and $\delta_{y} \geq$ and define rCI as:
\begin{equation}
    rCI = \frac{\text{\# Concordant Valid Pairs}}{\text{\# Valid Pairs}} = \frac{I(x_{i} > x_{j}, y_{i} > y_{j}, |x_{i} - x_{j}| > \delta_{x}, |y_{i} - y_{j}| > \delta_{y})}{I(|x_{i} - x_{j}| > \delta_{x}, |y_{i} - y_{j}| > \delta_{y})}
\end{equation}

The Kernelized Concordance Index (kCI) is a  generalization of rCI, where instead of only considering valid pairs, every pair is assigned a weight according to a kernel. The rCI is a special case of kCI with a heavyside kernel, where valid pairs are weighted with 1 and invalid pairs are weighted with 0. We choose the kernel so as the difference in measurements for either $\mathrm{x}$ or $\mathrm{y}$ approach 0, the weight for the pair of observations should tend to 0, and as the difference grows large, the weight should tend to 1. A sigmoid of the form $\frac{1}{1 + e^{kx - c}}$ is one such kernel with these properties, and we outline a procedure for fitting the kernel to problem-specific empirical data in the supplemental methods. The definition of the kCI formally is:
\begin{equation}
    kCI = \frac{2\sum_{i,j}w(\Delta x, \Delta y) I(x_{i} > x_{j}, y_{i} > y_{j})}{\sum_{i,j}w(\Delta x, \Delta y)}
\end{equation}
Where $\Delta x = |x_{i} - x_{j}|$ and $\Delta y = |y_{i} - y_{j}|$. It is not necessary to define the kernel purely on the difference in the observations, though we chose to do so for simplicity. 
For purposes of comparison, in subsequent analyses, we compared CI, rCI, and kCI with Pearson and Spearman correlation, which are also commonly used as test statistics for association. 

Both rCI and kCI have parameters that must be tuned for the intended data. The thresholds in rCI and the kernel shape in kCI can be interpreted as a characteristic range within which differences cannot be discriminated from statistical noise or measurement error. In the Methods, we define a method for choosing these parameters using problem-specific empirical data and replicate measurements. For our simulations below, we fit the parameters to real data from pharmacogenomic studies. 

\section*{Software}
To efficiently compute CI, rCI, and kCI, we developed an open-source R software package, wCI (weighted Concordance Index) that implements all three statistics and is freely available from

https://github.com/bhklab/wCI under the GPLv3. \\

The CI can be computed in $O(N \log N)$ time because counting the number of inversions in a permutation is computationally equivalent to sorting   \cite{knightComputerMethodCalculating1966}. 
Together with our work defining the rCI statistic, we present an algorithm for computing rCI also in $O(N \log N)$ time, described in a reference implementation included in the aforementioned repository. While the algorithm looks quite complex in implementation, intuitively it exploits three insights: The first is that the concordance index in can be computed in $O(N \log N)$ time by sorting both vectors by the ordering permutation for one, and then counting the number of "hops" each element in the other makes during a merge sort. The second is using "virtual" elements with values at $x-\delta_x$ and $x+\delta_x$, as well as $y-\delta_y$ and $y+\delta_y$, and computing the difference between the number of "real" elements hopped over during the merge sort for the $+\delta$ and $-\delta$ elements, with the difference between these two capturing the number of inverted pairs falling within the $2*\delta$ margin (deemed invalid pairs). The third is to repeat the algorithm twice, once starting with X sorted and once with Y sorted, while comparing the computed invalid pair counts and applying the inclusion/exclusion principle to compute the pairs deemed invalid during a single sort, or during both sorts. To keep the time complexity to $O(N\log N)$, we use a data structure to record the location of corresponding real and virtual items in $X$ and $Y$ and track their location when they are moved during the sort.\\

In its most general case, kCI can depend on a unique weight for each pair of elements. As such, it needs to read from memory $n \choose 2$ values, and therefore cannot be computed faster than $O(N^2)$. For certain restricted classes of kernels, faster computation may be possible, as evidenced by interpreting rCI as kCI with a Heavyside kernel, but an exploration of these special cases is outside the scope of our work. 

\section*{Results}

\section*{Inflation of P-values for Association Testing}
Assessing statistical significance requires comparing an observed value of the test statistic with a proper null distribution. The most commonly tested null hypothesis is one of no association, usually formalized as coefficient of association equaling zero for the population - this is the null hypothesis considered throughout this work. There exist analytical formulations for the distribution under the null hypothesis for all the coefficients above, except for kCI. For the Pearson correlation, the statistic $t=r\sqrt{\frac{n-2}{1-r^2}}$ is known to asymptotically follow a t-distribution with n-2 degrees of freedom, and is most commonly used in significance testing. For the CI, a widely applied analytical distribution of the concordance index, due to its general applicability to data including censoring and ties, was introduced by Noether \cite{noetherElementsNonparametricStatistics1967}, and advocated for application to the CI by Pencina and D’Agostino \cite{pencinaOverallMeasureDiscrimination2004}. The derived distribution is also only asymptotically correct for non-Gaussian data. For Spearman's rho, statistical testing is commonly done using an Edgeworth series expansion introduced by David et al. \cite{davidQUESTIONSDISTRIBUTIONTHEORY1951}.\\ 

\begin{figure}[H]
\centering
\begin{subfigure}{0.475\textwidth}
\centering
\includegraphics[width=\textwidth]{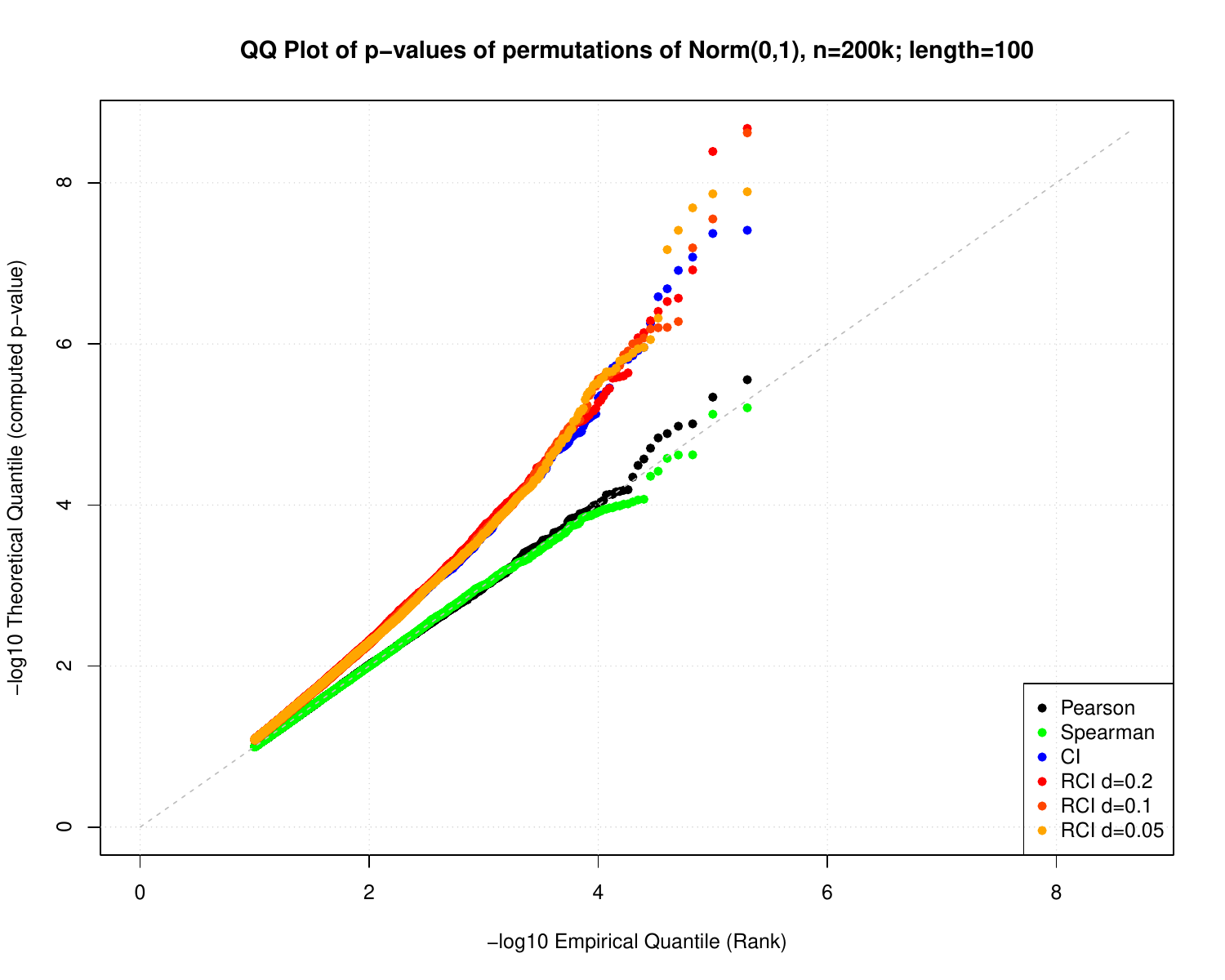}
 \caption{Fig1A}
 \label{fig:qqNormal}
 \end{subfigure}
 \hfill
 \begin{subfigure}{0.475\textwidth}
 \centering
 \includegraphics[width=\textwidth]{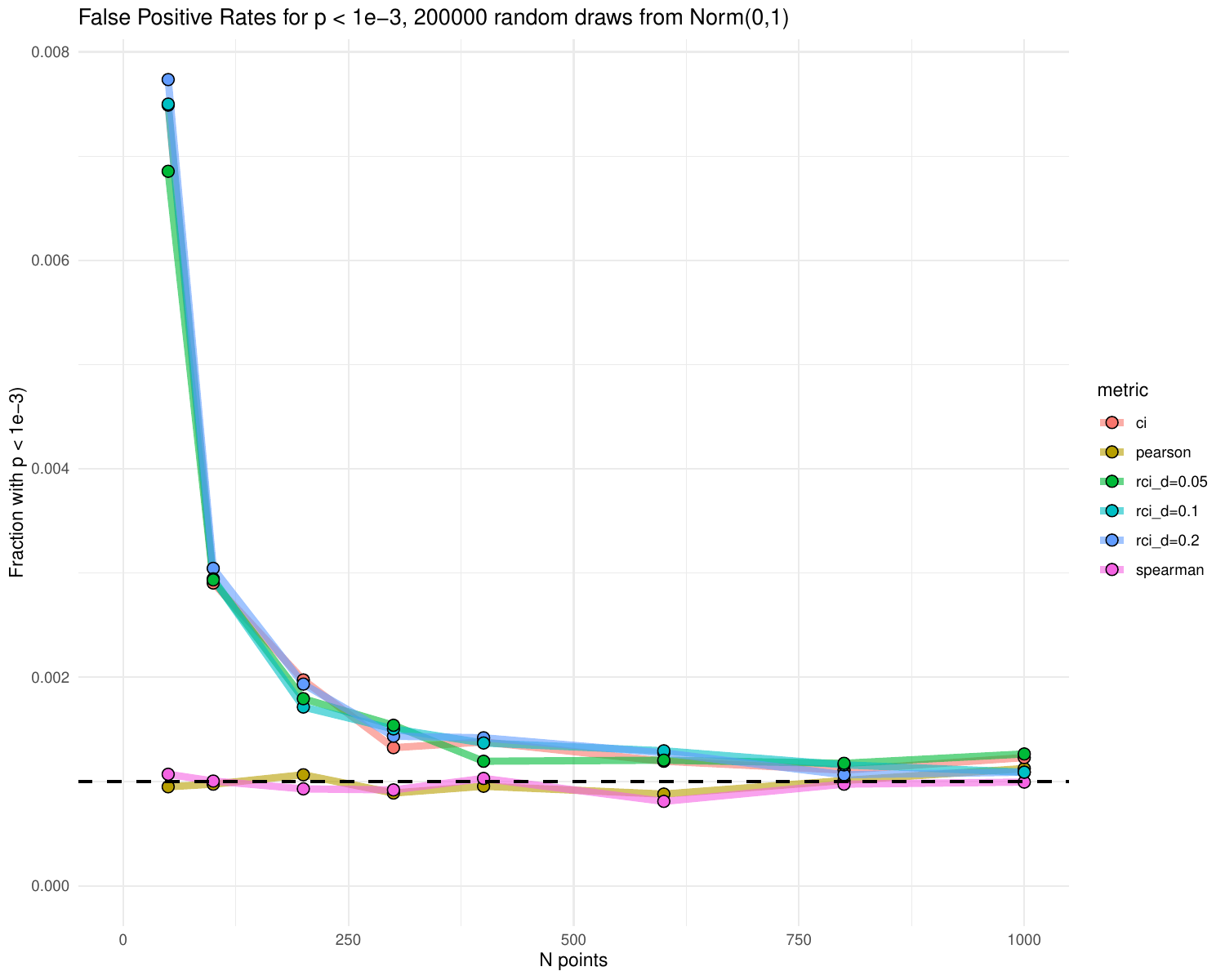}
 \caption{Fig1B}
 \label{fig:fprNormal}
 \end{subfigure}
 \vskip \baselineskip
 \begin{subfigure}{0.475\textwidth}
 \centering
 \includegraphics[width=\textwidth]{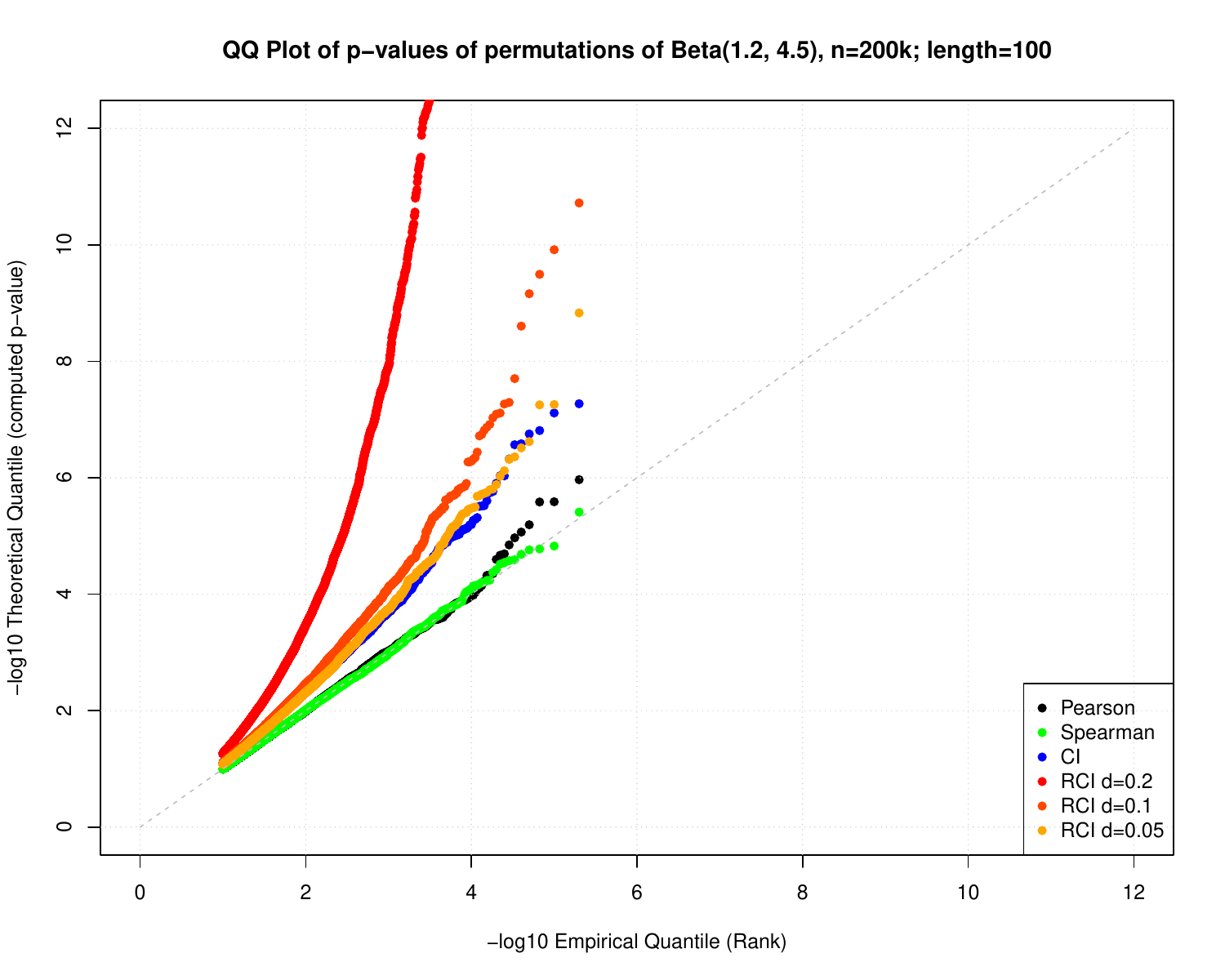}
 \caption{Fig1C}
 \label{fig:qqBeta}
 \end{subfigure}
 \hfill
 \begin{subfigure}{0.475\textwidth}
 \centering
 \includegraphics[width=\textwidth]{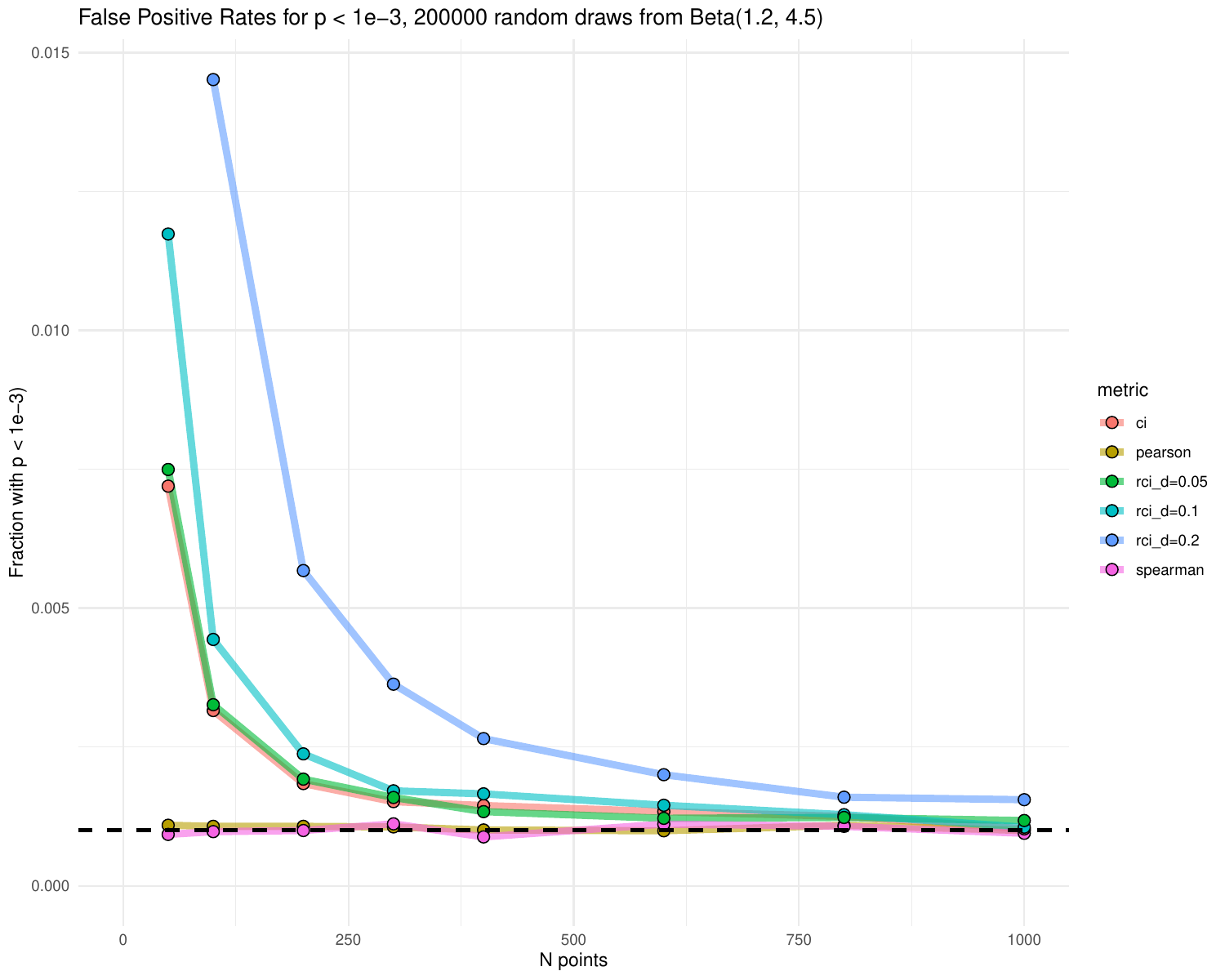}
 \caption{Fig1D}
 \label{fig:fprBeta}
 \end{subfigure}
 \caption{The asymptotic approximation of the CI null distribution produces an excess of small p-values. We took independent samples from a normal and beta distribution, computed their similarity using the coefficients above, and calculated asymptotic p-values using the approximations from the text. Because the samples are independent, their p-value distribution should be uniform. The QQ plots for normal (a) and beta (c) distributions for samples of length N = 100 sampled 200,000 times shows an excess of small p-values for CI and rCI. In the case of the normal distribution, p-values of $10^{-4}$ occur over twenty times more often than would be expected, and for the beta distribution nearly one hundred times more often for rCI. Figures (b, Normal) and (d, beta) summarize the frequency of $p < 10^{-3}$ for different sample sizes. As the number of samples grows large, the asymptotic approximation becomes more correct, but even in the regime of hundreds of samples, extreme p-values occur several times more often than they should under the null.}
\end{figure}

To evaluate the accuracy of these analytical formulations, we took independent samples from relevant distributions, computed the similarity according to each coefficient, and assessed the statistical significance using the analytical formulas. Under the null hypothesis, the p-values calculated from the analytical formula should be drawn from a uniform [0, 1] distribution. We summarize these findings with a Q-Q plot for both normally distributed data and beta-distributed data (Figures \ref{fig:qqNormal} and \ref{fig:qqBeta}, Supplementary figure \ref{fig:supplementalAllAAC}). In the context of testing association with a larger number $K$ of features - for instance, $10^{4}$ genes - multiple hypothesis testing requires the significance threshold be at least as small as the reciprocal of the number of features, $\alpha < \frac{1}{K}$. To illustrate this, we show the fraction of samples with analytical p-value less than several values of $\alpha$ (Figures \ref{fig:fprNormal} and \ref{fig:fprBeta}). \\

These results show that for the rCI and CI statistics, the analytical p-values are unreliable in the regime under 1000 observations, and that the problem is exacerbated for samples from non-normal distributions. This is also true to a smaller extent for the Pearson correlation test, in the Beta(1.2,4.5) case. The consequence of using the analytical p-values when the number of observations is small is an inflated type 1 error rate and false positive associations. 

\section*{Characterization of the Null Distribution of CI}
\begin{figure}[H]
\centering
\begin{subfigure}{0.475\textwidth}
\centering
\includegraphics[width=\textwidth]{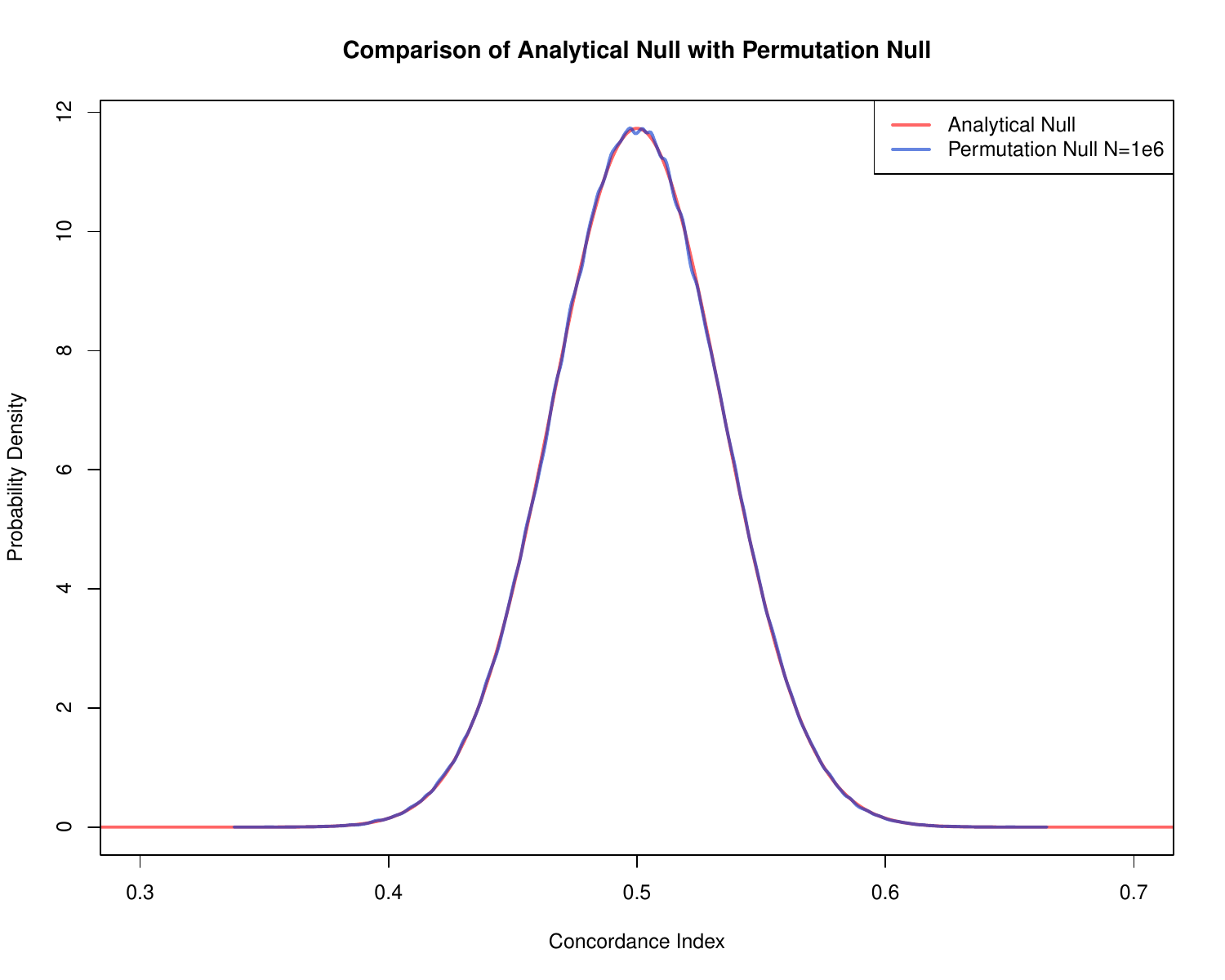}
\caption{Fig2A}
\label{fig:analVsPerm}
\end{subfigure}
\hfill
\begin{subfigure}{0.475\textwidth}
\centering
\includegraphics[width=\textwidth]{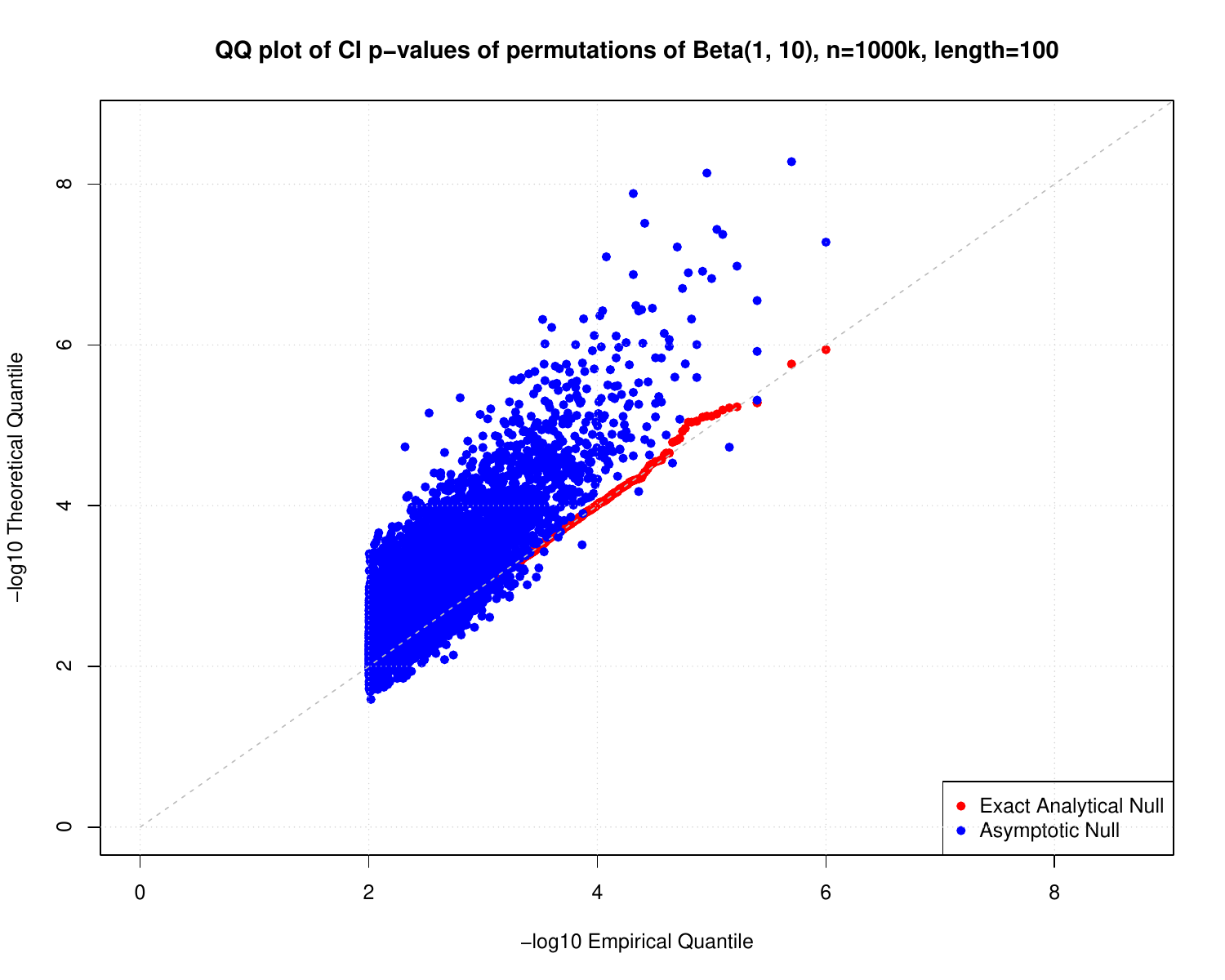}
\caption{Fig2B}
\label{fig:qqAnalNull}
\end{subfigure}
\caption{The analytical null accurately computes exact p-values for CI. (a) The analytical distribution matches a permutation null of K = 1e6 samples of length 100 from a standard normal distribution. As CI is entirely non-parametric, the choice of distribution is irrelevant. (b) The Q-Q plot shows the -$\log$10 empirical rank of the CI on the x-axis and the -$\log$10 theoretical quantile from the analytical null (red) and asymptotic null (blue). The analytical p-values are both monotonic and correctly approximate the uniform distribution (grey).}
\end{figure}

The null hypothesis for the CI is that the two variables have no association, and thus each permutation of the variables is equally likely. A permutation $\pi$ on n elements has an inversion if there exist two elements $i,j$ that are out of order, i.e. $\pi(i) > \pi(j)$ but $i < j$. The null distribution on CI is then the distribution on the number of inversions over the space of permutations of n elements. We have devised an exact null that can correctly compute the number of inversions on permutations of sets, in the absence of ties, of up to N = 170 elements. In cases of more than 170 elements, calculating the exact distribution breaks numerical precision of our environment, as the number of permutations goes as $N!$ and the range on the number of inversions is $C(N,2)$. The null distribution is also exact for pairs of variables which have a tie structure that can be represented as a multiset, i.e., observations can be partitioned into equivalence classes based on ties.
To demonstrate that the exact formula correctly calculates the null for CI with no ties, we computed the number of inversions from K = 1e6 permutations of N elements and compared this with our exact null. Comparing the exact distribution of the concordance index (1 - number of inversions divided by number of pairs) to the empirically observed distribution for the simulated permutations (Figure \ref{fig:analVsPerm}) we see very good consistency between simulation and our implementation of the exact calculation. We also computed the p-value according to our exact distribution for each simulation sample, and saw a very good concordance of the distribution of p-values with the uniform distribution, as expected under the null (Figure \ref{fig:qqAnalNull}). 

In general, if both variables have ties, the analytical null is unknown. This occurs with CI when both inputs have ties, with rCI in almost all circumstances, and for CI with right-censored data as in survival analysis contexts, all of which have a tie structure that cannot be represented by a multiset. Similarly, the exact analytical null for kCI is unknown. The only unbiased solution for computing statistical significance in cases where the limiting null approximation breaks down is a permutation null. Unfortunately, this is computationally costly. 

\section*{Power Analysis}
We set out to investigate the power of the newly defined rCI coefficient in comparison to the other commonly used correlation coefficients discussed above. As the theoretical distribution of the rCI statistic is unknown, we analyze the power of these statistics in simulation. We sampled from a simulated bivariate standard normal distribution with a known correlation between the two variables $r > 0$. For evaluating the power, we treat the population correlation as the effect size. We evaluated the statistical power of the Pearson, Spearman, CI and rCI coefficients for detecting an association at a fixed significance level $\alpha$, using a permutation-based test. In these simulations, we investigated the effects of effect size, sample size, and the rCI $\delta$ parameter on the power of these respective statistics. Lacking a principled way to choose a kernel for the kCI applied to normally distributed data in simulation, we did not include the kCI statistic in our analysis. \\

\begin{figure}[H]
\centering
\begin{subfigure}{0.5\textwidth}
\centering
\includegraphics[width=\linewidth]{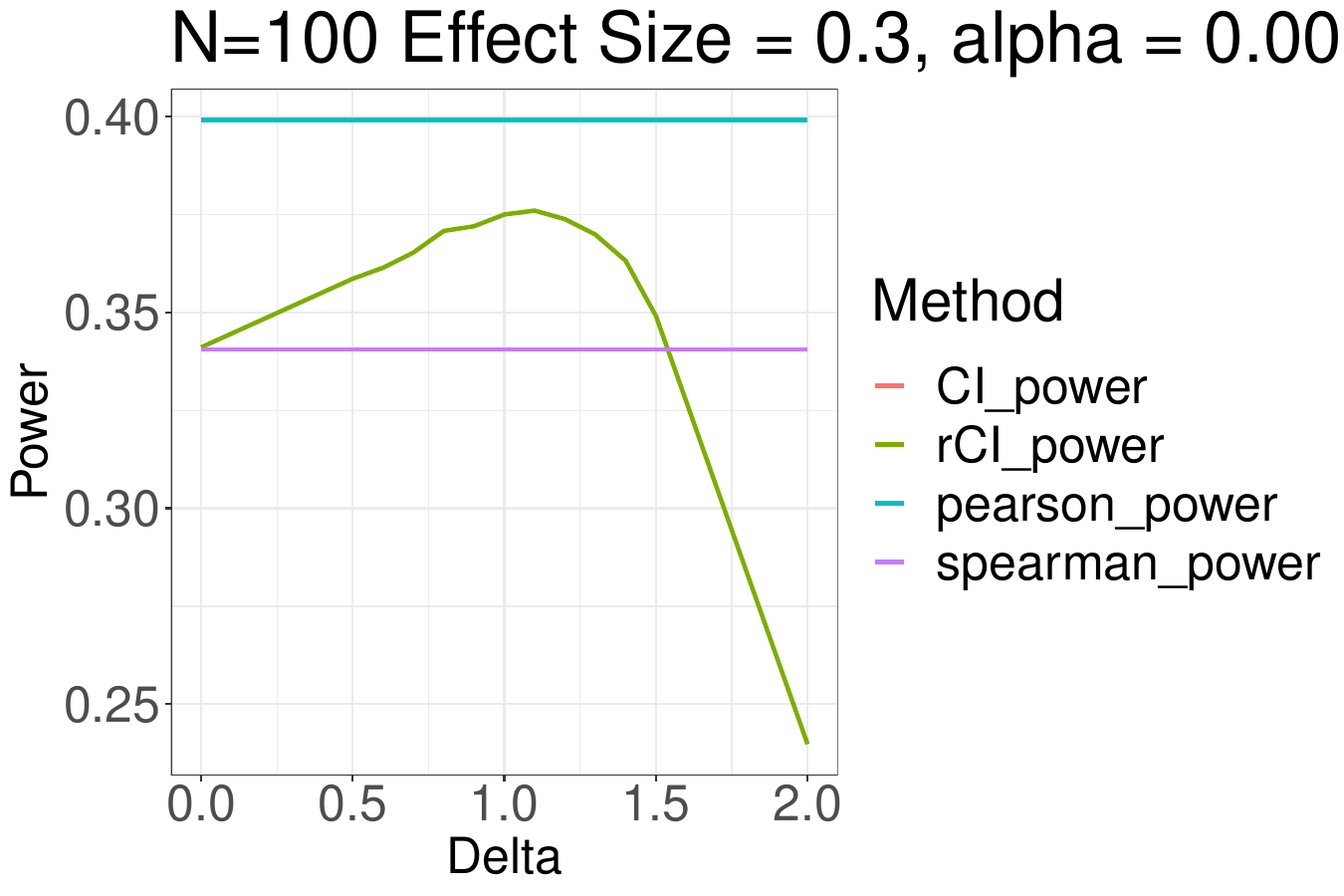}
  \caption{}
 \label{fig:gaussianPowerA} 
 \end{subfigure}~%
\begin{subfigure}{0.5\textwidth}
\centering
\includegraphics[width=\linewidth]{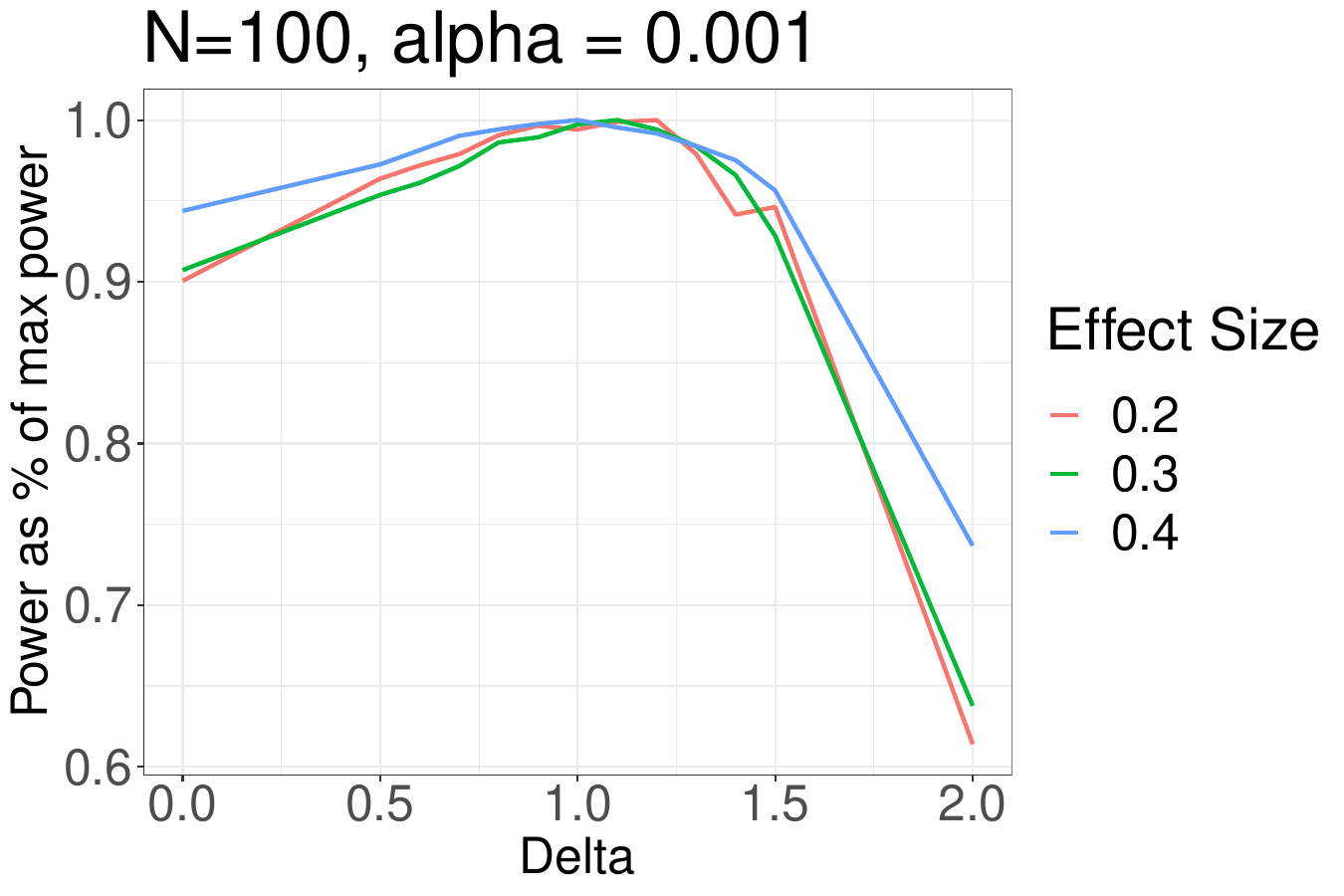}
\caption{}
   \label{fig:gaussianPowerB}
\end{subfigure}\\
\begin{subfigure}{0.5\textwidth}
\centering
\includegraphics[width=\linewidth]{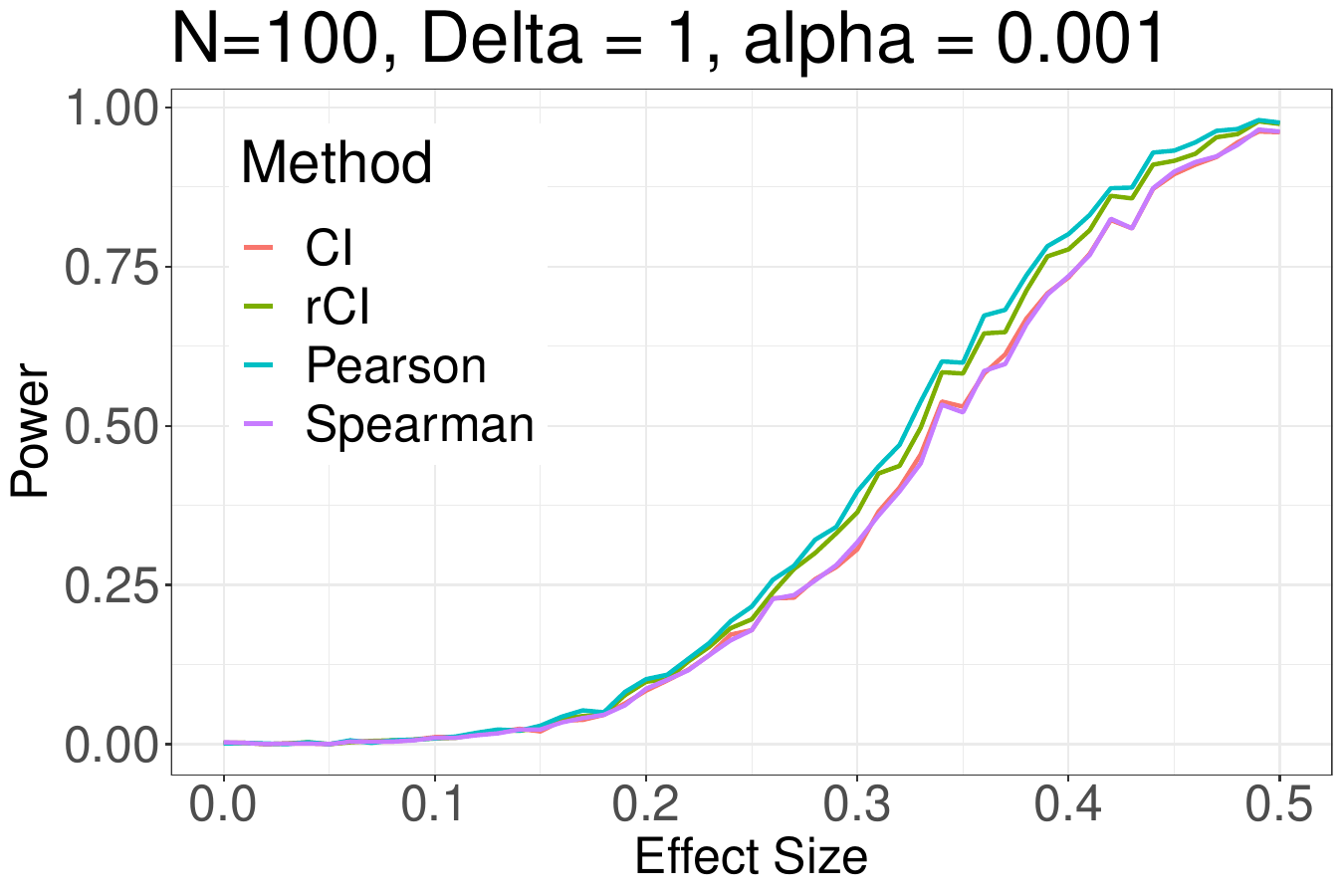}
 \caption{}
  \label{fig:gaussianPowerC}
\end{subfigure}~%
\begin{subfigure}{0.5\textwidth}
\centering
\includegraphics[width=\linewidth]{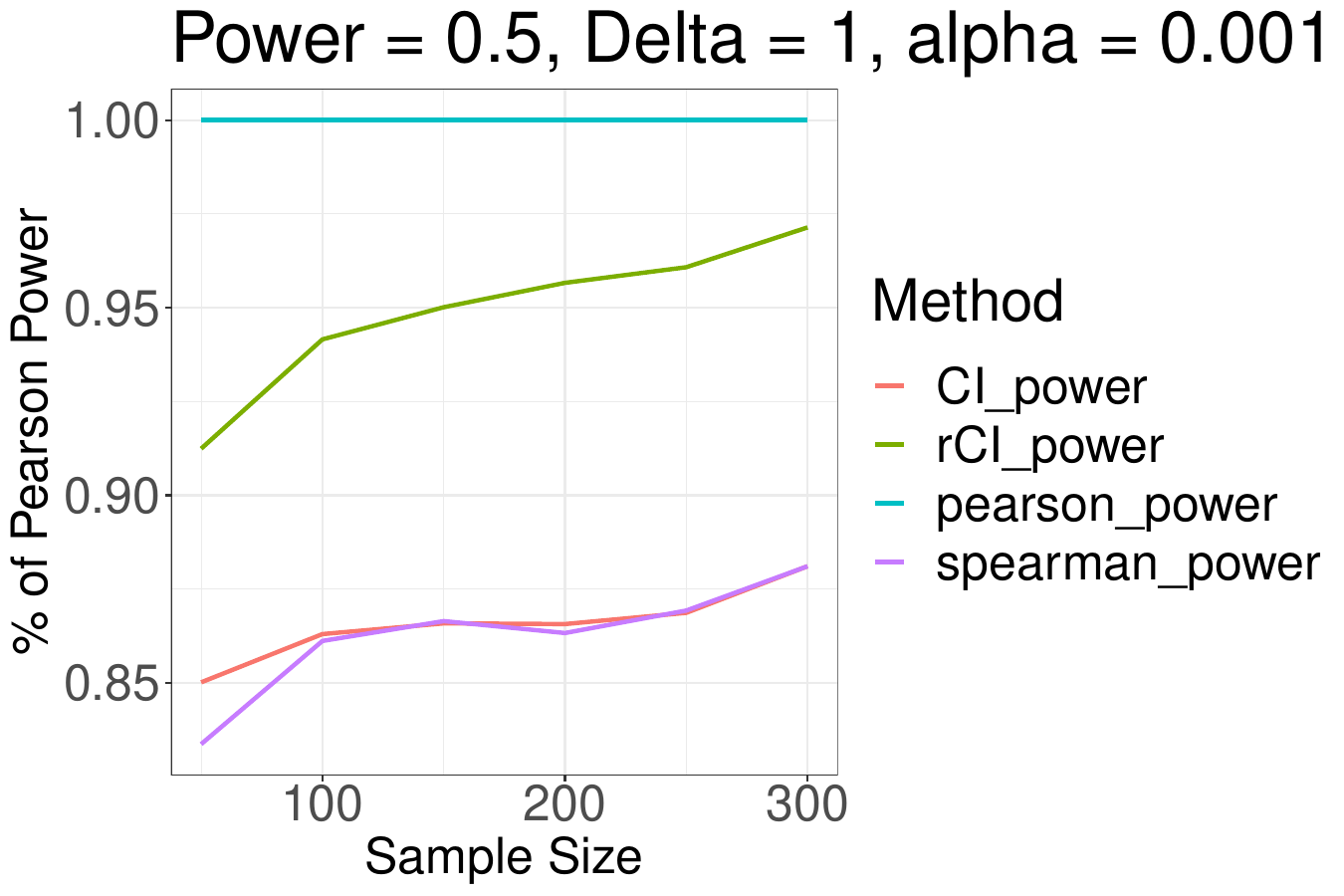}
 \caption{}
 \label{fig:gaussianPowerD}
\end{subfigure}
\caption{Power analysis for data simulated using the bivariate Gaussian family. a) displays the effect of the $\delta$ parameter on the empirical power at a fixed effect size of population r=0.3. Other statistics unaffected by the parameter are plotted for comparison. b) displays the empirically observed power for the rCI statistic only, plotting the dependence on delta at 3 different effect sizes. The power is normalized as percent of maximum power achieved for each effect size to highlight the optimal region for choosing delta. c) empirical power for as the population expected Pearson correlation increases. d) empirical power for a varying sample size, as the effect size is modified to keep a theoretically constant power for the Pearson correlation of 0.5. Power is plotted as the percent of achieved Pearson correlation power in simulation}
\end{figure}

We began by evaluating the effect of the new $\delta$ parameter on the power of the rCI statistic, seeking to answer the question of what choice of $\delta$ is optimal for this new statistic. We took $10,000$ samples of length $N=100$ from bivariate normal distributions with expected correlations of $0.2$, $0.3$ and $0.4$, representing low ($<10\%$), medium ($\sim 40\%$) and high ($>80\%$) power situations for the Pearson correlation with this sample size. Examining the medium power case ($r=0.3$), we see that the Spearman and CI statistics are similarly powered under this permutation test, whereas Pearson, as expected in this bivariate normal case, is significantly better powered than the most widely used non-parametric correlation coefficients (Figure \ref{fig:gaussianPowerA}). The rCI shows a trend to increasing power with increasing $\delta$ until approximately $\delta=1.2$, after which power decreases and at around $\delta=1.5$ the power of the rCI becomes worse than the CI or Spearman statistics. Note that in this case, the marginal standard deviation of the data is 1, so the $\delta$ parameter can be directly interpreted as a multiple of the standard deviation of the population. To compare the behaviour of the rCI statistic over the three different effect sizes investigated, we normalized the power at each $\delta$ to a percentage of the max power observed in simulation across all choices of $\delta$ (Figure \ref{fig:gaussianPowerB}). While the exact location of the maximum varies slightly with the effect size, the power is fairly stable in a range around $\delta=0.75$ and $\delta=1.25$ for all three investigated regimes, and close to the maximum achieved. Motivated by this observation, we chose $\delta=1$ for our subsequent simulations. \\

We then explored a range of expected population Pearson correlations between 0 and 0.5, looking at $1000$ samples of $N=100$ long vectors for each effect size and computing the empirical power observed in simulation (Figure \ref{fig:gaussianPowerC}). Unsurprisingly, the Pearson statistic is most powerful across the range, followed by rCI (at $\delta=1$), and then Spearman and CI practically indistinguishable. These simulations reinforce the benefit of the rCI modification over existing fully non-parametric statistics across the full range from weakly powered to well powered situations. \\

Finally, we wished to investigate how the sample size on which the correlation coefficients are calculate affects their power in a permutation test. For this, we decided to calculate a level-set of constant power as decreasing effect size is traded off for increasing sample size, fixing the expected Pearson power to be at 50\%, and investigating sample sizes from $N=50$ to $N=300$, in steps of $50$. For this simulation, we did $10,000$ samples from the distribution to calculate the power. We then normalized the power observed at each sample size as percentage of observed Pearson power to adjust for slight errors in calculation of the level set (Figure \ref{fig:gaussianPowerD}). We see a clear trend among rCI and the non-parametric statistics of increased power compared to the Pearson correlation as sample size increases, however, while rCI achieves $>90\%$ of the Pearson power already at $N=50$ and $>95\%$ by $N=300$, the non-parametric statistics remain at less than $90\%$ power even at $N=300$.

\subsection*{Power Simulation in Non-Normal Data}

Our ultimate motivation for investigating the performance of existing correlation coefficients and proposing modifications of the concordance index was their application to noisy biological data. Prior to examining real data, we sought to evaluate the power of these permutation tests in cases where the marginal distributions of the observed variables is non-normal. Motivated by our intended application of drug screening data, we chose to investigate data drawn from bivariate beta distributions. In particular, we simulated data drawn from distributions with both marginals fixed as $\mathtt{Beta}(1.2,4.5)$ distributions (Supplementary figure \ref{fig:supplementalBetaDistPlot}), and varying Pearson correlations between the two sampled variables. Further details on the simulation are described in the methods section. \\

Unlike the normally distributed data, we did not optimize the selection of delta parameters for rCI through power simulations. Rather, we derived these parameters from real pharmacogenomics data, by choosing a threshold on differences between AAC measurements that maximizes our classification performance of replicate vs non-replicate measurements (see Methods). The optimal threshold was very close to 0.1, so for the simulations, a delta of 0.1 was used. We also included the kCI statistic in these simulations, deriving kernel parameters from the same data (see Methods). \\

We first examined the empirical power of these statistics at a modest sample size of 100 samples, varying the correlation between the two variables. We again explored a range of expected  Pearson correlations between 0 and 0.5, looking at $1000$ samples of $N=100$ long vectors at each effect size. While the Pearson correlation was more powerful at weaker effect sizes, at larger effect sizes (where all statistics achieve $>50\%$ power in simulation), the CI and kCI statistic become slightly more powerful in detecting significant effects at this sample size (Supplementary Figure \ref{fig:supplementalBetaPower1245}). The rCI statistic performs similarly to Spearman over the range of effect sizes investigated, both falling short of the Pearson correlation until all statistics converge to 100\% power. \\

We then investigated how sample size affects the power of these statistics. Similar to our previous simulations with normally distributed data, we used a theoretical calculation of a level set of 50\% power for the Pearson correlation test. With the data distribution deviating significantly from normal, this was even more approximate than before, so we once again normalized everything relative to the power we observed for the Pearson correlation in simulation. We again took 10,000 samples each for vectors of length 100-300, in step sizes of 50. We observe that as sample size increases larger than 100, the power of the CI statistics in a permutation test continues to increase over the Pearson correlation (Figure~\ref{fig:supplementalBetaPowerLevelset}). The kCI statistic behaves similarly to the Pearson correlation in these simulations, while the rCI and Spearman, both gaining power relative to the Pearson correlation, but do not equalize in power even in our largest sample size tested, 300. We repeated the same simulations, but adjusting effect size to hold a theoretical power of 0.25 (Supplementary Figure~\ref{fig:supplementalBetaPowerLevelset25}). Consistent with our previous results, at a sample size of 100 the Pearson correlation was the most powerful statistic, but as the sample size grew the CI overtook the Pearson correlation in power, and it was the most powerful statistic at sample sizes of 300. The rest of the statistics increased in power similarly as before, but starting off relatively less powerful compared to Pearson. \\ 

\subsection*{Power Simulation in Non-Normal Data With Additive Noise}

Finally, we repeated the simulations at fixed sample size and varying expected correlations with Beta distributed data above, with the addition of noise sampled from a Laplace distribution, mimicking noise seen between replicate experiments in AAC measurements, as detailed in the methods section. The additive noise was significantly smaller in magnitude than the range of the data, and as the rCI $\delta$ and kCI kernels were picked using the replicate measurements, we expected these statistics to be advantaged in terms of power to detect effects "corrupted" by noise. \\

All statistics were less powerful after the addition of noise (Figure~\ref{fig:supplementalBetaPowerNoise}). This is unsurprising, as from the point of view of a sample statistic, the data generation process is opaque: a particular sample generated with added noise is indistinguishable from data generated with a weaker between-variable correlation. In these simulations, the rCI and kCI exhibit a similar drop in power to non-parametric statistics. Interestingly, the Pearson correlation was actually less affected by the addition of noise than all the other statistics, leading to the Pearson correlation outperforming all other statistics in observed power throughout the range of effect sizes simulated. It is possible that these results are observed because the addition of additive noise increases slightly the variance of the data, and may decrease the skewness of the distribution, reducing the deviation from normally distributed data.

\section*{Querying Pharmacogenomic datasets}
\begin{figure}[H]
\centering
\begin{subfigure}{0.475\textwidth}
\centering
\includegraphics[width=\textwidth]{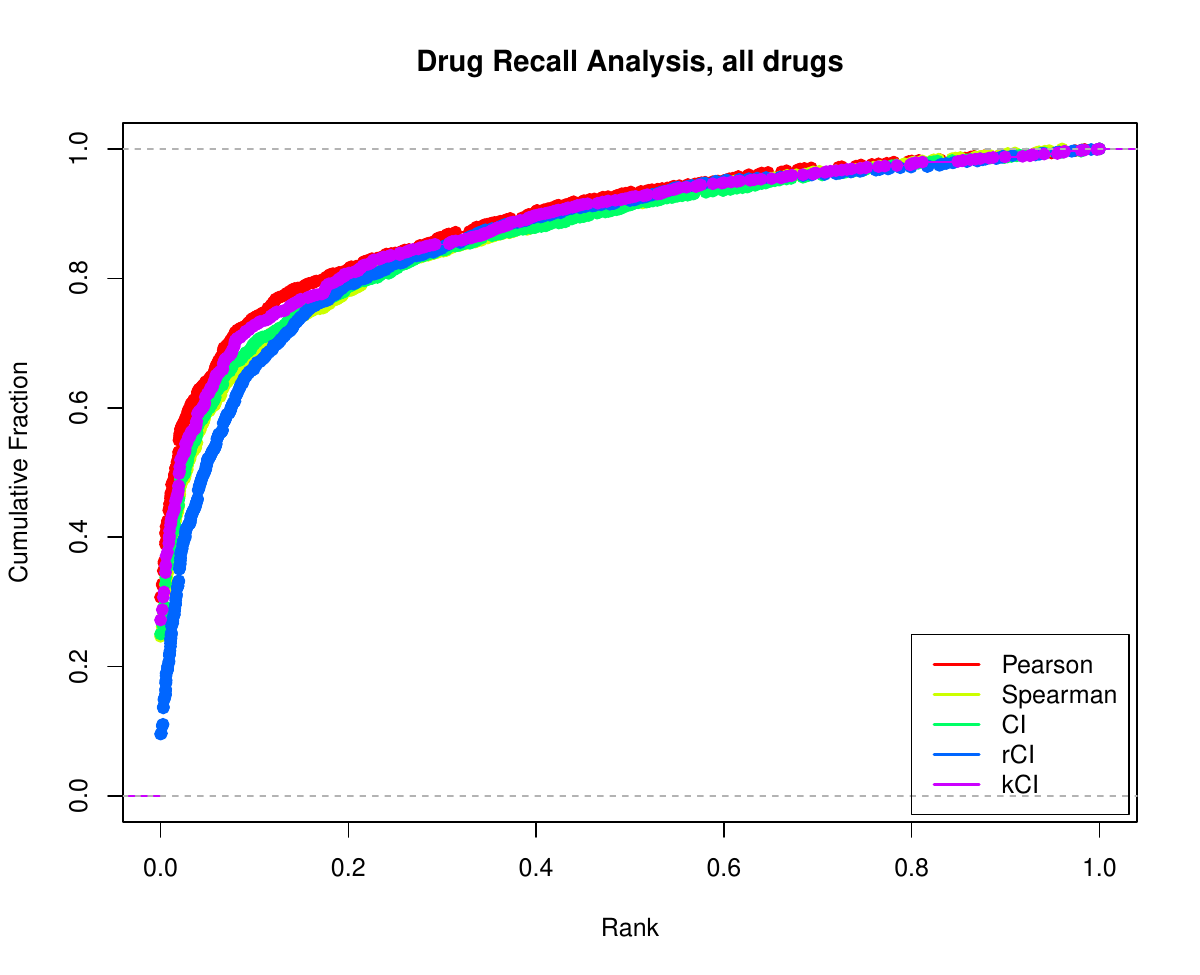}
\caption{}
\label{fig:drugrecall}
\end{subfigure}
\hfill
\begin{subfigure}{0.475\textwidth}
\centering
\includegraphics[width=\textwidth]{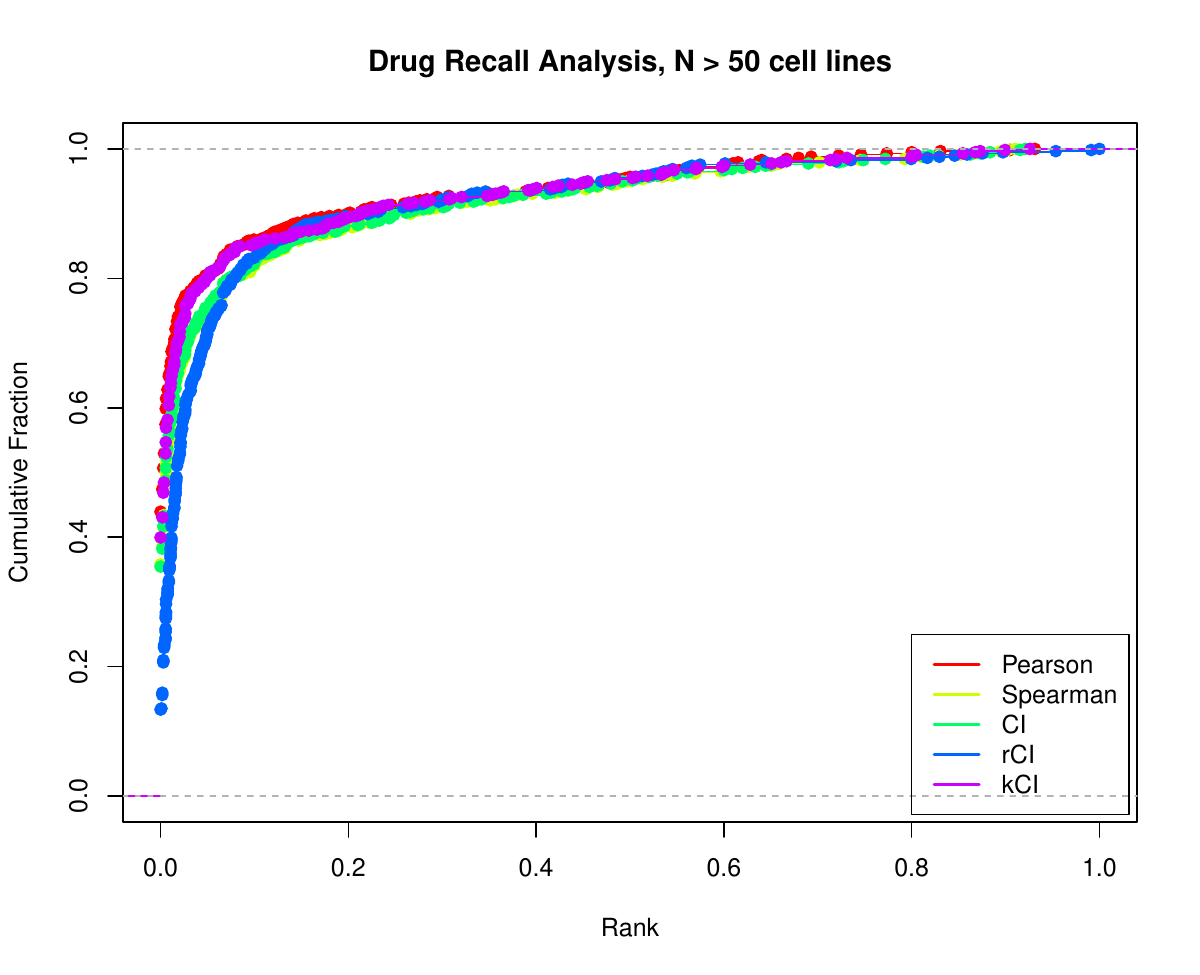}
\caption{}
\label{fig:drugrecall50}
\end{subfigure}
\caption{Drug recall analysis across pharmacogenomic datasets. For all pairs of datasets, the similarity between the vector of cell line responses for all pairs of drugs is computed with each coefficient for (a) all drugs and (b) those drugs with at least fifty cell lines in common across datasets. For drugs present in both datasets, the rank of the matched drug relative to all drugs is extracted. The x-axis is the rank of the matched drug, where 0 is most similar and 1 is least similar. The y-axis is the empirical CDF of the matched drugs for a given rank, or the fraction of matched drugs with rank less than x. }
\end{figure}

With the theoretical properties explored, we sought to evaluate the practical performance of these coefficients on real pharmacogenomic data. Large-scale pharmacogenomic studies have treated panels of cancer cell lines with drugs at a range of doses. The sensitivity of a cell line to a drug is summarized with the area above the dose response curve (AAC), and assaying a panel of cell lines gives a vector of drug sensitivities for a particular drug. For each drug measured in two studies across a common set of cell lines, we computed the similarity coefficients between that drug's response vector in one study against all other drugs' response vectors in the other study. We then found the rank of the matched drug (Figure~\ref{fig:drugrecall}). This has been shown to be a challenging problem in that drug sensitivities do not necessarily replicate well across pharmacogenomic studies \cite{Haibe-Kains2013InconsistencyStudies, Haverty2016ReproduciblePanels}. The statistic with best resilience to the noise and artifacts in the pharmacogenomic studies will tend to rank matched drugs better. Because some of the drugs have very few cell lines intersected across datasets, we also considered only those drugs with more than 50 cell lines (Figure~\ref{fig:drugrecall50}). \\

The five metrics had similar performance, though Pearson had the consistently best recall. The kCI showed improved recall across datasets compared to CI and Spearman correlation, suggesting that incorporating information characterizing the noise in the datasets can improve identification of association. While this benchmark is informative, it is ultimately inconclusive about the practical capabilities of the different coefficients to identify biomarkers of drug sensitivity. 


\begin{table}[h]
\centering
\begin{tabular}{c | c | c}
Coefficient & Area under CDF & Area under CDF, N $>$ 50 cell lines\\
\hline
Pearson & 0.8897 & 0.9395\\
Spearman & 0.8702 & 0.9238\\
CI & 0.8694 & 0.9248 \\
rCI & 0.8639 & 0.9238 \\
kCI & 0.8805 & 0.9354
\end{tabular}
\end{table}


\section*{Discussion}

Evaluation and significance testing of coefficients of association is a ubiquitous step in modern biological data analyses. However, the behaviour of common correlation coefficients applied to noisy data originating from distributions seen in modern biological data is not well characterized. Our study addressed this question for high-throughput cell line viability screening. We proposed new measures of association for noisy data, investigated the correctness of significance testing for common correlation coefficients, introduced efficient and accurate implementations for calculating significance for the concordance index, and investigated the power of permutation testing coefficients of association in distributions mimicking our area of application. Finally, we demonstrated that incorporating information about the measurement noise into calculating associations between vectors of observations, through our proposed kCI approach, demonstrates improved performance over the standard concordance index in a cross-dataset replicability task. \\ 

Researchers in biology will often compute and test thousands of correlations, requiring multiple testing correction to keep the expected number of false positives under control, often done through a False Discovery Rate or Family Wise Error Rate correction procedure. Hypothesis testing in these situations require methods that can compute p-values accurately in the range of 1e-4 - 1e-10, depending on the number of hypotheses investigated, meaning that correct tail end behaviour dominates the accuracy of inference. Pencina and D'Agostino \cite{pencinaOverallMeasureDiscrimination2004} investigated the use of Noether's method to calculate 95\% confidence intervals in the presence of complicated tie structures due to right censoring of one vector, and showed the method to have moderately precise coverage. Unfortunately, our data suggests that the extension of this approach to calculating p values for hypothesis testing does not control false positives at lower alphas than 0.05, at moderate to large sample sizes. This is the case not only for the rCI statistic which introduces tied pairs in the data, but also for the concordance index applied to data with no ties. While this may be unsurprising given the asymptotic arguments for the derivation of Noether's method, what caught us off guard in our results was the slow rate at which the control of false positives converged, even for the case without ties and sampling from a bivariate normal distribution. As much as 500 samples were not enough to restore proper false positive control at an alpha of 0.001, and given the behaviour observed in the Q-Q plots for smaller sample sizes, we expect the sample size required for more stringent alpha's will be substantially larger. \\

Permutation tests are a versatile tool for assessing significance without making strong parametric assumptions about the generating distribution for the data. For permutation testing to be theoretically justified in making inferences about a population parameter, the data has to be independently and identically sampled, and each observation has to satisfy exchangeability between the two variables measured \parencite{hayesPermutationTestNot19960101}. These assumptions are all shared by both exact tests for non-parametric correlations, and asymptotic analytical tests for the Pearson and CI correlations. Therefore, from a theoretical basis, the permutation test can only increase the range of valid situations for its application. Two concerns often raised by practitioners regarding applying permutation testing is a perceived loss of power, and the computational cost of simulating permutation null distributions. It is important to note that an inflation of false positives will lead in practice to lower p values and the illusion of more power originating from analytical tests. Previous studies suggest however, that the power of the permutation test for the Pearson correlation is very nearly identical to the Pearson t test in all cases where the t test retains good type 1 error control \parencite{hayesPermutationTestNot19960101, Bishara2012TestingApproaches}, with the t test providing significantly more power only in cases where data is nearly normally distributed, and at low (\~5) sample sizes \parencite{Bishara2012TestingApproaches}. Concerns about the computational cost of permutation testing are valid, especially in the context of multiple testing correction. However, approaches such as QUICKSTOP \parencite{heckerFlexibleNearlyOptimal2020}, for cases where the required alpha after correction is known, as well as other approaches focused on ensuring FDR control \cite{guoAdaptiveChoiceNumber2008}, can reduce by orders of magnitude the number of permutations needed to perform inference. Combined with modern computer hardware, permutation testing, even at "high-throughput data scale", becomes feasible. Therefore, based on our results assessing Type I error rates for asymptotic approximations of the CI distribution, and prior work mentioned above, we advocate for permutation testing to be adopted as the default testing strategy for tests of association, and require strong justifications on reverting to analytical formulas for calculating p values for correlation coefficients. \\

Once it has been established that permutation testing is necessary to obtain correct p value estimates for sample sizes of interest, we compared the power of the different statistics under the permutation test. Pearson's correlation coefficient remained the most powerful statistic to detect linear associations between normally distributed variables when using a permutation test, but with only a marginal advantage. The power difference between the Pearson, CI, rCI and Spearman are small, and probably of little significance in practice at our investigated sample sizes and $\alpha$. The newly defined rCI statistic exhibited most power with the delta parameter in the range of approximately 1 standard deviation of the data, and the value of the optimal delta was fairly stable over different effect sizes. While less powerful than the Pearson correlation in this simulation, the semi-parametric rCI was more powerful than non-parametric statistics, suggesting that in cases where the Pearson is not a suitable statistic (for example, looking for non-linear relationships), the rCI could be the tool of choice.  \\

Beyond the Gaussian case, none of Pearson, Spearman or CI are consistently most powerful over the range of effect sizes investigated in bivariate beta distributed data at a moderate sample size of 100. The performance of our newly defined statistics, rCI and kCI, was similar to the others. Interestingly, we noticed that the Concordance Index seemed to outperform the Pearson correlation in the well powered ($>50\%$) regime. Strikingly, holding the power of the Pearson test at 50\% while increasing sample size, the relative power gain of the Concordance Index increases. This suggests that for standard regimes of biological data analysis, where data is moderately large (100s of samples), and non-normally distributed, a permutation test based on the Concordance Index may be better powered to detect significant effects than the widely used Pearson correlation. We must caution however that precise rankings of these correlations in distributions very different from the skewed and bounded data we investigated cannot be extrapolated from our results. \\

There are, of course, limitations to this work. First, we only investigated the task of detecting linear associations from data. Undeniably, non-linear effects play a major role in biology, however linear analysis is often the first step in understanding any biological system. 
A major focus of our work has been on the concordance index (and by extension the Kendall Tau) measure of correlation. We have however omitted any investigation of the CI (and our derivatives) to censored data, where it is most commonly applied. Finally, due to computational constraints, we could not exhaustively search the space of distributions, sample sizes, hyper-parameters, and effect strengths. \\

In conclusion, we urge researchers analysing biological data to take care in choosing both a coefficient for measuring association within their data, and a corresponding test for statistical significance. Examining the distribution of p-values returned by the chosen statistical test under the null hypothesis should be a standard step in statistical data analysis, and analysts should make sure to investigate $\alpha$ levels relevant to their application and multiple testing burden. Furthermore, when applying coefficients measuring association to biological data, non-parametric statistics may exhibit greater statistical power than parametric ones, especially in larger sample sizes. Therefore, even a humble a task as calculating correlations between observed variables requires a careful, critical and open-minded approach from a modern biological scientist.   \\

\section*{Methods}

\subsection*{Cell Line Screening Datasets Considered}

As mentioned previously, our investigations were motivated by the task of finding correlations between high throughput compound screens measuring viability in cancer cell lines. For this study, we considered data originating from six different screening datasets: the Genomics of Drug Sensitivity in Cancer (versions 1 and 2) \cite{iorioLandscapePharmacogenomicInteractions2016,Yang2013GenomicsCells.,garnettSystematicIdentificationGenomic2012}, the Cancer Therapeutics Response Portal v2 \cite{seashore-ludlowHarnessingConnectivityLargeScale2015, reesCorrelatingChemicalSensitivity2016, Basu2013AnMolecules}, the Genentech Cell Screening Initiative \cite{havertyReproduciblePharmacogenomicProfiling2016}, the Finnish Institute of Medicine cell line screening dataset \cite{mpindiConsistencyDrugResponse2016} and the Oregon Health and Science University Breast Cancer Screen \cite{hafnerQuantificationSensitivityResistance2017}. All the data used in this study was previously published, except for the data from the Genentech Cell Screening Initiative, where an updated version of the original dataset was used for our analyses.  \\

All datasets were downloaded using the PharmacoGx R package \cite{Smirnov2016PharmacoGx:Datasets}, from orcestra.ca \cite{mammolitiOrchestratingSharingLarge2021}. For the analyses in this manuscript, all dose-response curves were summarized as the Area Above the Curve, using a 3 parameter Hill Curve model fit using the PharmacoGx package, as previously described \cite{safikhaniRevisitingInconsistencyLarge2017}. 

\subsection*{Characterization of the Null Distributions}
There are four situations to consider:
\begin{enumerate}
    \item Concordance Index with no ties - the two vectors are composed of unique elements, i.e. $\nexists i,j$ such that $x_{i} = x_{j}$ or $y_{i} = y_{j}$
    \item Concordance Index with ties - $\exists i,j$ such that $x_{i} = x_{j}$ or $y_{i} = y_{j}$. This problem breaks down further into two cases: (A) the tie structure is transitive and can be represented by one multiset - if $i \cong j$ and $j \cong k$, then $i \cong k$, or (B) the tie structure cannot be represented by a multiset, so $\exists i,j,k | i \cong j, j \cong k, i \ncong k$, where $\cong$ denotes a tie in at least one of the vectors. 
    \item Robust Concordance Index - we define two thresholds $t_{x}$ and $t_{y}$ such that if $|x_{i} - x{j}| < t_{x}$ or $|y_{i} - y_{j}| < t_{y}$, the pair is considered invalid and ignored in the CI calculation.  The denominator in the count of inversions is the number of valid pairs, where both $\Delta x$ and $\Delta y$ exceed their respective thresholds. As with non-transitive tied CI, rCI's tie structure does not necessarily form an equivalence relation and is not transitive. 
    \item Kernelized Concordance Index assigns a weight $w$ to all pairs that is a function of one or both of the differences between the values.  Larger differences in x and y can be considered a pair with greater confidence.  The choice of kernel is typically monotone increasing with delta, for instance a sigmoid.  rCI can be thought of as a special case of kCI where the kernel is a Heavyside step function.
\end{enumerate}

The exact analytical nulls for 1 and 2A are known and presented here. For cases 2B, 3, and 4, the exact null is unknown, so statistical significance must be assessed with permutation testing. \\ 

\subsubsection*{Case 1}

In the cases that there are no ties, calculating the concordance index can be reduced to counting inversions in a permutation. Given two lists of numbers without ties, $x$ and $y$, to calculate $CI(x,y)$, it is sufficient to find the ordering permutations $\sigma_x$ and $\sigma_y$ for $x$ and $y$ respectively, and then count the number of inversions in $\sigma_x \cdot \sigma_y^{-1}$. Computationally, this is equivalent to ordering both lists such that $x$ is sorted, and then counting inversions in y. \\

The simplest way to generate the probability distribution on the number of inversions from a set is to use a generating polynomial or an ordinary generating function. The distribution on the permutations which result in a particular inversion number is represented as coefficients in a polynomial on x; the number (or probability, if normalized) of permutations yielding k inversions is the coefficient of $x^{k}$. \\

In the absence of ties, from \cite{Margolius2001}, the following result holds: let $I_n(k)$ denote the number of permutations of $S$ with $k$ inversions.  Then:
\begin{equation}
        \Phi_n(x) := \sum_{j=1}^{n(n-1)/2} I_n(x) x^k = \prod_{j=1}^n \sum_{k=1}^{j-1} x^k.
        \label{eq:noTieInv}
\end{equation}

The right hand side of \eqref{eq:noTieInv} admits exact recursive and iterative methods for adding elements by multiplying polynomials.  This is equivalent to convolutions on discrete probability distributions. We note that this formula is widely used in practice, for example, the R Statistical language \cite{rcoreteamLanguageEnvironmentStatistical2020} implements this exact formula for calculating the significance of Kendall's Tau correlation for small sample sizes (by default 50 at the time of writing) in the absence of ties, testing against the null hypothesis of no association. \\  

\subsubsection*{Case 2a}
 
The best way to think about permuting elements with ties between them is as follows: Each equivalence class formed by the ties in the original vectors $x$ and $y$ is assigned to a specific element $e_j$. Let the set of distinct elements to be permuted be denoted $E=\{e_1,...,e_n\}$ and let $a_j\in \mathbb{Z}_{>0}$ denote the multiplicity of element $e_j$, i.e. how many elements are in this equivalence class. Thus there are $\alpha:=\sum_{j=1}^n \alpha_j$ elements in total. We denote by $M=\{e_{1}^{\alpha_{1}},...,e_{n}^{\alpha_n}\}$ the multiset containing all elements (with ties).\\

It turns out that an analogous result to \eqref{eq:noTieInv} can be obtained for multisets. The original reference is from 1915 -- see \cite{macmahonCombinatoryAnalysisVolumes2001} -- but it's easier to read in modern notation, such as presented in \cite{Remmel2015}. Let $\inv(\sigma)$ denote the number of inversions of a permutation of the multiset (set with ties) $M$. The \emph{distribution} of $\inv$ can be shown to be
\begin{align}
\label{eq:mnc}
    D_M(x) &= \sum_{\sigma\in S_M} x^{\inv(\sigma)} = \begin{bmatrix} \alpha\\ \alpha_1 ... \alpha_n \end{bmatrix}_x = \frac{\alpha!_x}{\alpha_1!_x .. \alpha_n!_x}
    \intertext{with the \emph{$q$-factorial} being defined by}
    \label{eq:DefQFac}
    m!_x & = \prod_{k=1}^r \left(1+x+...+x^{k-1}\right)
\end{align}
(The expression on the right-hand side of \eqref{eq:mnc} is also called the \emph{$q$-multinomial coefficient}.) Observe that, by splitting the sum over $S_M$ according to the number of inversions,
\begin{align}
    D_M(x) = \sum_{k\geq 0} \sum_{\substack{\sigma\in S_M\\ \inv(\sigma)=k}} x^{\inv(\sigma)} = \sum_{k\geq 0} \sum_{\substack{\sigma\in S_M\\ \inv(\sigma)=k}} x^k = \sum_{k\geq 0} I_M(k) x^k
\end{align}
where $I_M(k)$ denotes the number of permutations of the multiset $M$ with $k$ inversions. Thus, \eqref{eq:mnc} is the exact multiset analogue of \eqref{eq:noTieInv}.\\

The right hand side of \eqref{eq:noTieInv}, gives a recursive formula for expressing $I_n(k)$ in terms of the $I_{n-1}(j)$: in terms of the generating function this reads
\begin{align}
\label{eq:IterativeNoties}
    \Phi_n(x) &= \left(\sum_{k=0}^{n-1} x^k\right) \Phi_{n-1}(x).
\end{align}
The proof if this formula (presented in \cite{Margolius2001})  proceeds by looking at permutations of the first $n-1$ elements and then inserting the last element at all possible position. By keeping track of how many extra inversions this insertion introduces, we arrive at \eqref{eq:IterativeNoties}. 

This argument extends rather well to the case with ties: let $M$ be the multiset as described in the introduction and denote by $M^-$ the set obtained from $M$ by removing one occurrence of $e_n$. That is, if $M=e_1^{\alpha_1},..,e_n^{\alpha_n}$, then
\begin{align}
    M^- = e_1^{\alpha_1}, e_2^{\alpha_2},\ldots,e_{n-1}^{\alpha_{n-1}}, e_n^{\alpha_n-1},
\end{align}
and in particular if $\abs{M}=n$ then $\abs{M^-}=n-1$. We can give the following analogue of \eqref{eq:IterativeNoties}:
\begin{align}
 D_M(x) & = \begin{bmatrix}\alpha\\ \alpha_n\end{bmatrix}_x D_{M-}(x) = \frac{\alpha!_x}{(\alpha-\alpha_n)!_x \alpha_n!_x} D_{M-}(x),
 \label{eq:IterativeTies}
\end{align}
with $m!_x$ defined in \eqref{eq:DefQFac}.\\

We note that this case, where the Concordance Index can be reduced to counting inversions on multisets is of practical interest. Specifically, it will arise whenever only one of $x$ or $y$ contains exact ties. However, it is limited to exact ties, as representing the vectors as a multiset requires transitivity of ties, and therefore cannot be applied to censored data. \\

\textbf{Notes on efficient implementation:}

Calculating the distribution of inversions over random permutations requires polynomial multiplication, or equivalently, convolution of discrete distributions. This can be accomplished in $O(n \log(n))$ time using the Fast Fourier Transform; once transformed, polynomial multiplication is elementwise instead of convolutional.  Numerical precision limitations introduce very slightly complex coefficients after the inverse transform introduces, which can be dealt with by rounding to the nearest integer. The naive application of FFT for polynomial multiplication would be to take the two factors, transform them, multiply, and inverse transform them.  We implemented a method by which the sequence of polynomials to multiply is generated, transformed and multiplied simultaneously, and inverse transformed once to further optimize the process. \\

Computing the distribution on inversions for Case 2a requires dividing polynomials. It turns out that in all cases encountered when applying the recursive formula \eqref{eq:IterativeTies} - with the $D_{n}$ ratios of $\prod_{k=n}^{n+m}z_{k}$ to $\prod_{k=1}^{m+1} z_{k}$, for each polynomial in the denominator, there always exists a polynomial in the numerator which is a polynomial multiple of the denominator such that the coefficients are integral and the remainder is 0. In our experience, doing any polynomial division - including in FFT space - was prohibitively inaccurate, so we implemented a method to simplify the $D_{n}$s to just the product of the quotients, eliminating the division entirely.\\

Efficient implementations of these distributions in R are shared with the wCI R package. 

\subsubsection*{Cases 2b - 4}

We note that the exact distribution under the assumption of no association between the two vectors $x$ and $y$ for Cases 2b - 4 are not known. For Cases 2b, as well as 3, the multiset representation breaks down. It is likely that a completely different approach will be needed to efficiently (in polynomial time) compute these general cases. Note that the CI applied to data with censoring falls into this regime. \\

The continuous nature of the kCI kernel likewise suggests that the null distribution can not be computed combinatorically. If solved however, the distribution for rCI will likely be a limiting case as the kernel approaches the discontinuous Heavyside function.

\subsection*{Power Analysis}
All power analysis was conducted through simulations in the R programming language. A link to the code is provided at the end of the manuscript. 

Each power analysis simulation followed the same steps:
\begin{enumerate}
\item Two vectors of data of chosen length N were sampled from a bivariate distribution with a known expected Pearson correlation $R$ as the effect size measure.
\item A permutation test was conducted using each of the Pearson, Spearman, CI, rCI and kCI (for Beta distributed data) as a test statistic, using the adaptive permutation testing algorithm described below. Tests resulting in P values under the chosen $\alpha$ of 0.001 were considered to have rejected the Null hypothesis, and therefore successfully detected the dependence between the two vectors. 
\item Steps 1-2 were repeated 1000 or 10,000 times (as described in the results section), recording for each iteration and each statistic the whether the Null was rejected. The power at each effect size was calculated as the percentage of cases in which the Null was rejected by each statistic.
\item Steps 1-3 were repeated for effect sizes ranging from $R=0$, to $R=0.5$, in $0.1$ step increments.
\end{enumerate}

Bivariate Normal Data was generated using the function \texttt{mvrnorm} from the \texttt{MASS} R package.  

\subsubsection*{Simulating Data Distributed as a Bivariate Beta }

We construct a bivariate Beta distribution with a fixed expected Pearson correlation from the family of distributions described in \cite{olkinConstructionsBivariateBeta2014}. Briefly, the distributions are constructed by sampling $(U_1, U_2, U_3)$ from a trivariate Dirichlet distribution, and transforming the variables as follows: $X = U_1 + U_3$, $Y = U_2 + U_3$. This construction is a 4 parameter family, with 3 parameters fixed by the shapes of the marginal distributions of $X$ and $Y$. The fourth parameter is then optimized to yield the desired correlation between $X$ and $Y$, using the Brent method as implemented by the R \text{optim} function. 

\subsection*{Adaptive Permutation Testing}

Adaptive stopping during permutation testing was used for all power analyses. Briefly, adaptive stopping allows the permutation test to complete once there are enough random resamples to determine whether a particular sample falls above or below a predefined significance threshold $\alpha$. For these analyses, the QUICK-STOP algorithm presented by \cite{heckerFlexibleNearlyOptimal2020} was implemented in the R programming language. The parameters used for our simulation were as follows: the indifference parameter $d$ was set to $0.001*\alpha$, the Type I and II error probability was set to $e^{-10}$, and a stopping criteria was implemented to halt after $100*1/\alpha$ permutations. 

\subsection*{Choice of rCI thresholds and kCI Kernel}

The Robust and Kernelized Concordance indices have parameters that can be optimized for the particular distribution of interest: the rCI threshold and the kCI kernel. Because both coefficients are premised on downweighting pairs of points with differences that cannot be distinguished from statistical noise, replicate measurements are needed to characterize the noise distribution from the data. Conceptually, the goal is to identify pairs of points with true measurements that are different given the random variable of the delta between the measurements. The approach is similar to Storey's FDR \cite{Storey2002ARates}. \\

Let $x$ be a quantity for a set of N points, and $\hat{x}$ be the measurement of x. For a pair of points $i$ and $j$, consider $\Delta \equiv |x_{i} - x_{j}|$. The null hypothesis $h_{0}$ is that the quantity is the same for the pair, i.e. $x_{i} = x_{j}$. The alternate hypothesis is that the two values are different: $x_{i} \neq x_{j}$. Let $S$ be a set of empirical deltas where the true difference is unknown. For repeated measurements of the same point, the null hypothesis is necessarily true, which gives the distribution of $\delta$ under they null hypothesis. Let $S_{0}$ be a set of empirical deltas from replicates. The problem is then to deconvolve $S$ into a mixture of null and alternate hypotheses. \\

The main result uses the empirical CDF on $S$ and $S_{0}$. From Bayes' Theorem, we have the probability of a pair of points having equal values given a measured delta exceeding some value t as:
\begin{equation}
P(h_{0} | \Delta > t) = \frac{P(\Delta > t | h_{0}) P(h_{0})}{P(\Delta > t)} \leq \frac{P(\Delta > t | h_{0})}{P(\Delta > t)}
\end{equation}
Here, the numerator is the ECDF of $\Delta$ for replicates multiplied by the estimate of the fraction of observed points for which the null hypothesis is true, and the denominator is the ECDF of $\Delta$ for the population of measurements. The inequality emerges because $P(h_{0}) \leq 1$. Similarly, the probability of the alternate hypothesis can be computed from above by the law of total probability, that $P(h_{0}) + P(h_{1}) = 1$. \\

We consider deltas exceeding the threshold t as positive cases, and those with deltas less than t as negative cases. From this, a confusion matrix can be generated as a function of the threshold:
\begin{center}
\begin{tabular}{l|l|c|c|c}
\multicolumn{2}{c}{}&\multicolumn{2}{c}{True Value}&\\
\cline{3-4}
\multicolumn{2}{c|}{}&Positive&Negative \\
\cline{2-4}
rCI decision boundary & Positive & $P(h_{1} | \Delta > t)$ & $P(h_{0} | \Delta > t)$ \\
\cline{2-4}
& Negative & $P(h_{1} | \Delta \leq t)$ & $P(h_{0} | \Delta \leq t)$ \\
\cline{2-4}
\end{tabular}
\end{center}

The threshold for rCI is given by choosing $t=\tau$ that maximizes the Matthew's Correlation Coefficient (MCC) from the confusion matrix above. While a contingency table cannot be perfectly summarized by any one value, MCC has been shown to be a balanced measure of classification even when the class sizes are different. For the kCI kernel, it is desirable to have a monotonic kernel that is 0 at $\Delta = 0$ and without loss of generality tends to 1 as $\Delta$ gets large. The $P(h_{1} | \Delta > t)$ fits these criteria well, with the additional stipulation at 0. For portability, we fit a sigmoid function to $P(h_{1}|\Delta > t)$, and this is the kernel: the weight assigned to a pair of points is the estimate of the probability that they have different true values. 

The thresholds or kernels for both rCI and kCI depend only on $\Delta$ and not on the specific values of the measurements. This implicitly assumes homoscedasticity of $\Delta$ over the range of the measurements. While it might be useful to explore this further, this choice was motivated by practicality, as there often is insufficient data to characterize the heteroscedasticity of $\Delta$.


\subsection*{Simulation of Artificial Noise}

To simulate noise representative of real pharmacological screening data, we first examined the distribution of pairwise differences in AAC values for experimental replicates done by the same study across all of the datasets described above. Upon observing this data, it was clear that it was not normally distributed. Using Maximum Likelihood Estimation, we fit a Gaussian [$a \exp{(\frac{-(x-b)^2}{2c^{2}})}$] and Laplace distribution [$\frac{1}{2b}\exp{(- \frac{\abs{x - \mu}}{b})}$] to the data, and found the Laplace fit to be qualitatively good, and adequate for use in a simulation context (Figure \ref{fig:supplementalLaplaceNoisePlot}). \\

For power simulations sampling from the Beta distribution with artificially added noise, two vectors of correlated data with a chosen expected correlation were sampled as above. A vector of the same length was then sampled from the MLE Laplace noise distribution, and added to one of the two vectors of simulated data. Finally, any values under 0 or over 1 after the addition of noise were truncated to keep the data bounded in the correct range. 

\subsection*{Research reproducibility}
The wCI package and analysis code is available on github, the pharmacogenomic datasets used in the analysis is available via the orcestra.ca platform \cite{Mammoliti2020ORCESTRA:Analyses}, and the code and data is accessible in a Code Ocean capsule. 

\subsection*{Querying Pharmacogenomic datasets}

To gauge the performance of the coefficients on practical pharmacogenomic data, we considered the four pharmacogenomic datasets mentioned above obtained from PharmacoGx. We compared two datasets at a time, and compared the matrix of drug sensitivities summarized with the area above the dose response curve as described above. First, the pair of matrices were restricted to those cell lines common to both studies. Then, for each drug in the first study, the similarity to all drugs in the second study was computed using each of the coefficients. For those drugs assayed in both studies, the rank of the similarity for the matched drugs was found relative to the similarity of the drug to all drugs in the second study. This self-rank of a drug was computed over all pairs of the pharmacogenomic datasets. We summarized the ranks using the area under the CDF of the ranks. An area under the CDF of the ranks of 1 indicates that the most similar drug in the second study to that of the first was the matched drug in all cases. If the matched drug ranks were uniformly distributed - indicating no correspondence between studies, then the area under the curve would be 0.5. \\

An obvious caveat is that the intersections of cell lines across studies was often small. If few common cell lines were measured, then the drug was poorly characterized and the comparison was less meaningful. To mitigate this, in \ref{fig:drugrecall50}, we considered only those drugs with at least 50 cell lines in common across datasets. This improved the performance of all the coefficients. It is also worth considering the performance of the different metrics for different regimes of drugs, i.e. those with targeted effects with few sensitive cell lines contrasted with those with broad effects and many sensitive cell lines. The parameters for the rCI and kCI were chosen based on combined replicates from all studies; the threshold for rCI was set to 0.12, and the kCI kernel had parameters (-27.52, 0.0646). 

\section*{Acknowledgements}
We thank Richard Bourgon for reviewing an early draft of this work, the constructive feedback provided made this work stronger and more rigorous. 
\section*{Supplemental Figures}
\renewcommand{\thefigure}{S\arabic{figure}}

\setcounter{figure}{0}

\begin{figure}[H]
    \centering
    \includegraphics{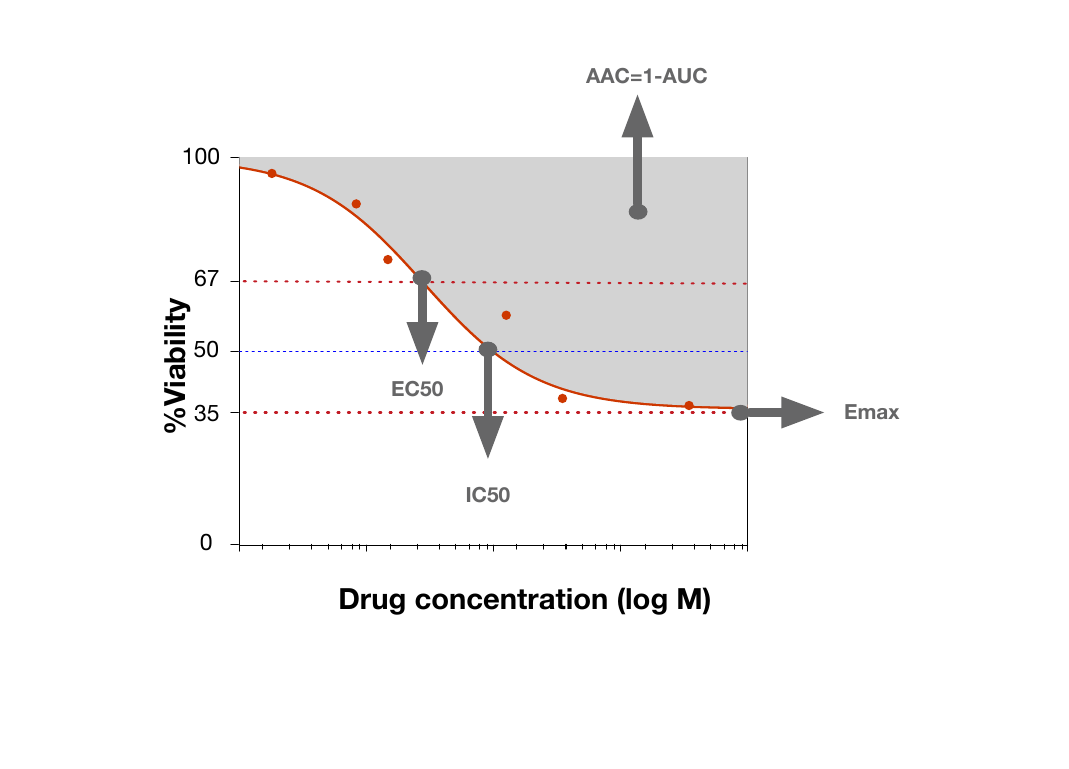}
    \caption{A prototypical drug dose - viability curve, including a representation of various metrics which can be calculated to summarize the curve behaviour. Reproduced from Safikhani et al. \cite{safikhaniChapterPharmacologicalGenetic2016}. }
    \label{fig:ddrcMetrics}
\end{figure}

\begin{figure}[H]
\centering
\includegraphics[width=\linewidth]{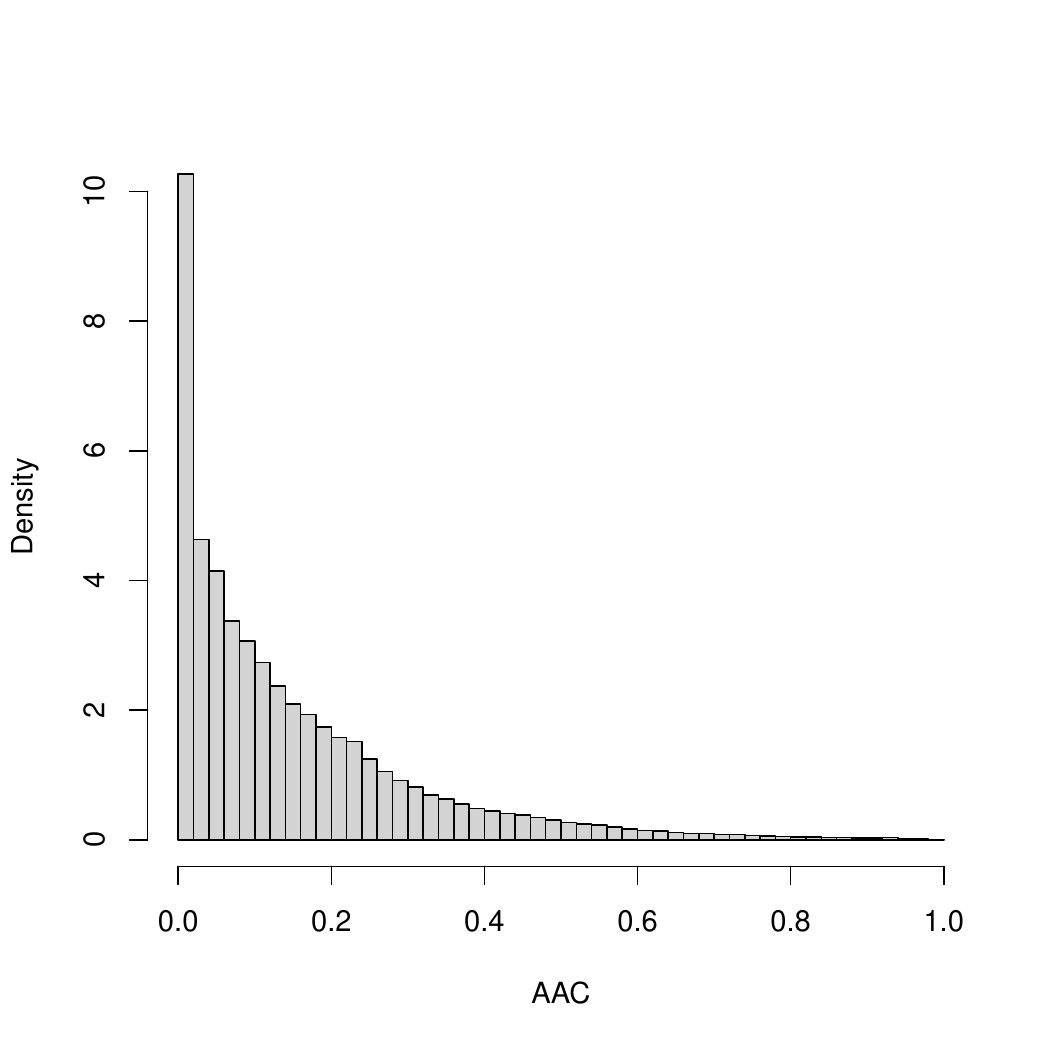}
 \caption{The distribution of all AAC values investigated in this study, combining across cell lines, drugs and studies. We can see the bounded and skewed behaviour which is simulated by a Beta distribution in our work.}
 \label{fig:supplementalAllAAC}
\end{figure}

\begin{figure}[H]
\begin{subfigure}{0.4\textwidth}
\centering
\includegraphics[width=\linewidth]{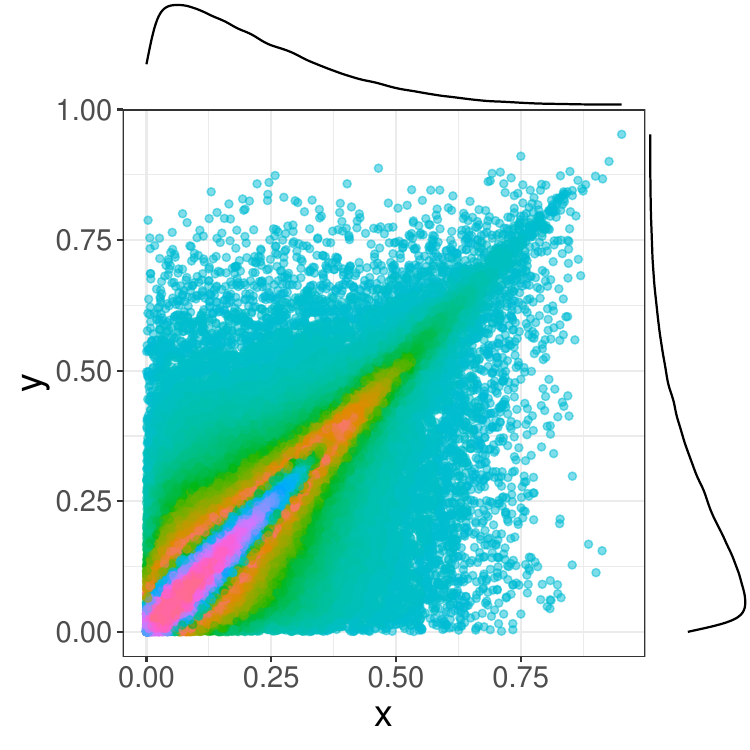}
 \caption{}
 \label{fig:supplementalBetaDistPlot}
\end{subfigure}~
\begin{subfigure}{0.6\textwidth}
\centering
\includegraphics[width=\linewidth]{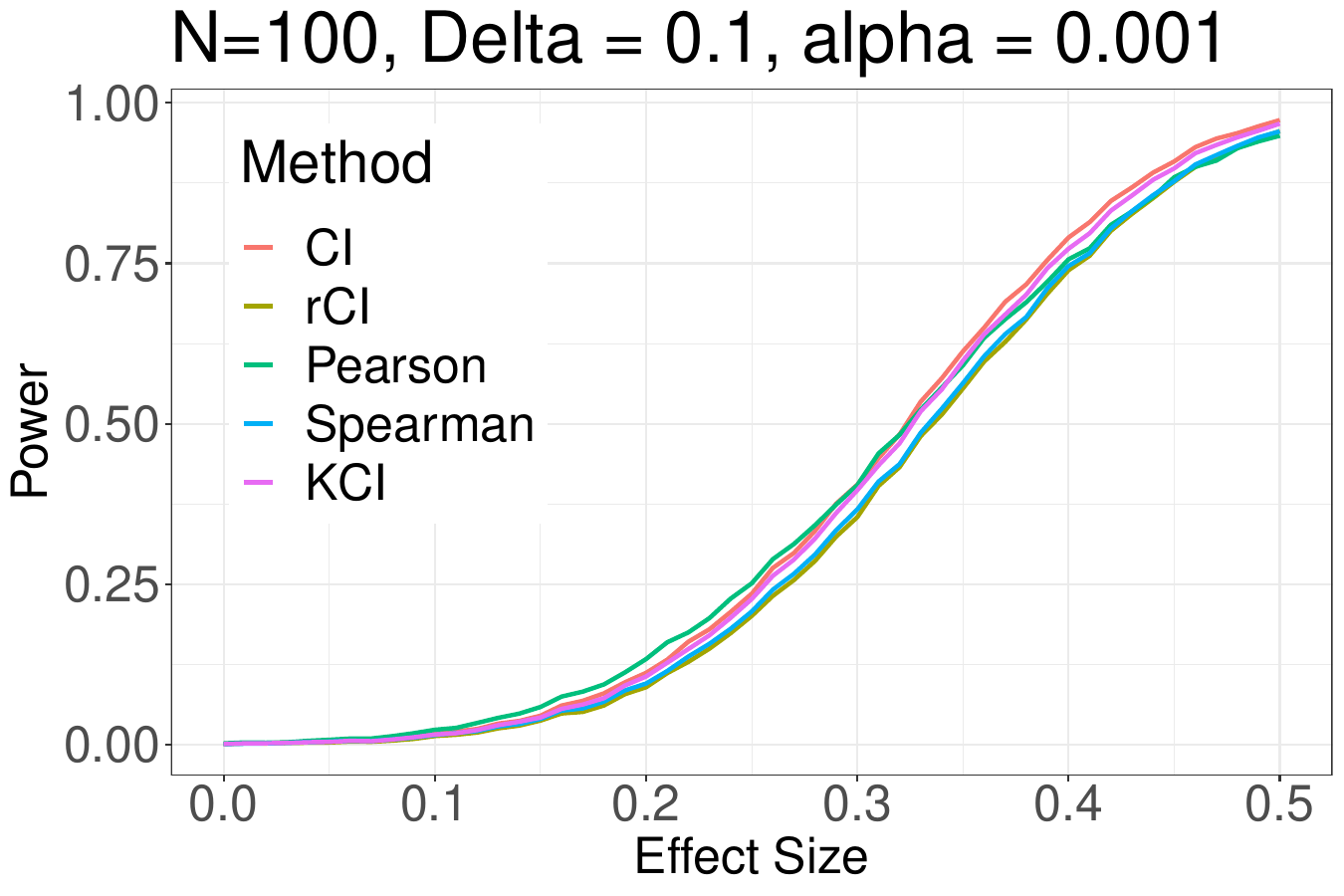}
 \caption{}
 \label{fig:supplementalBetaPower1245}
\end{subfigure}\\

\begin{subfigure}{0.5\textwidth}
\centering
\includegraphics[width=\linewidth]{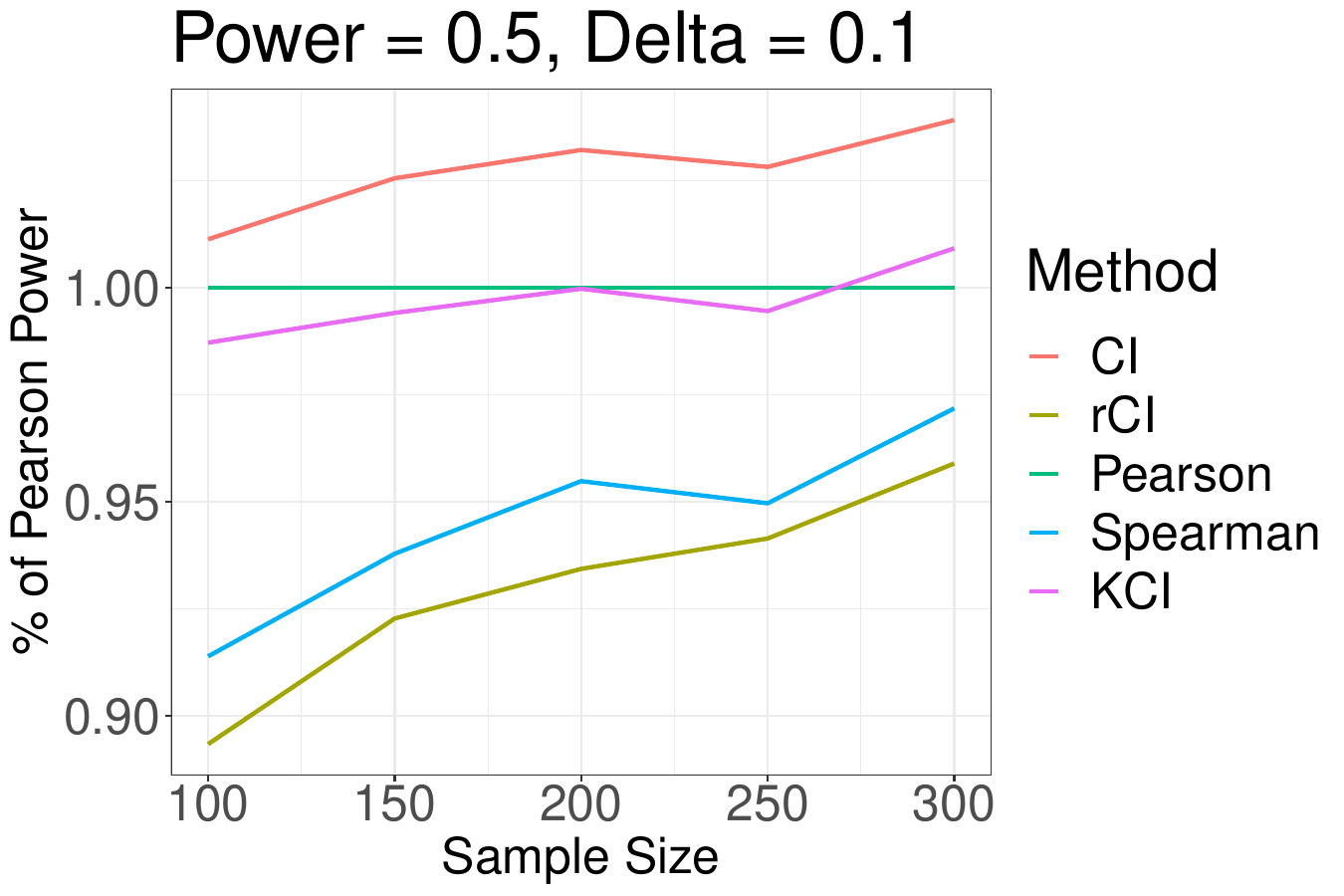}
 \caption{}
 \label{fig:supplementalBetaPowerLevelset}
\end{subfigure}
\begin{subfigure}{0.5\textwidth}
\centering
\includegraphics[width=\linewidth]{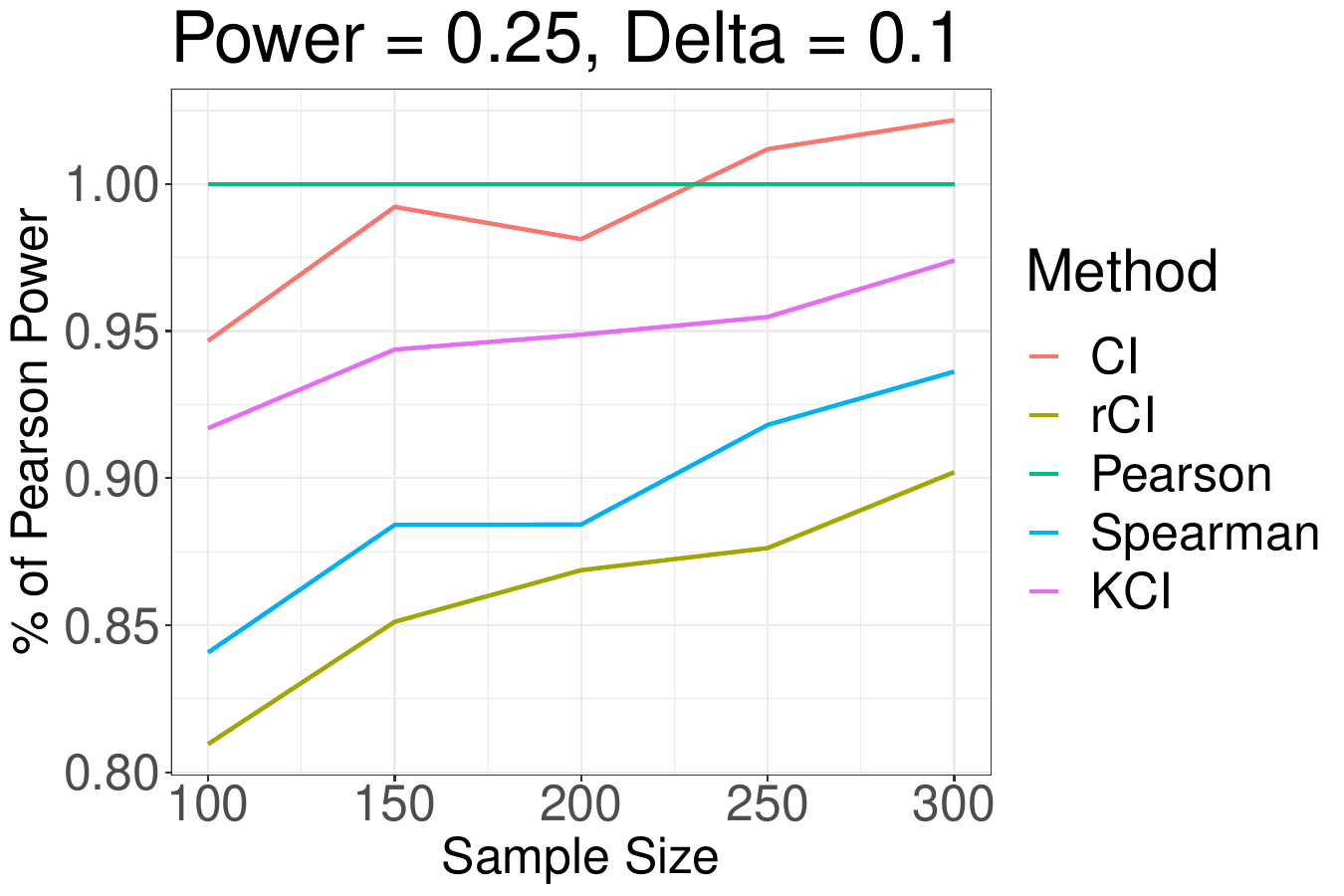}
 \caption{}
 \label{fig:supplementalBetaPowerLevelset25}
\end{subfigure}
\caption{Results of power analysis simulating Beta(1.2,4.5) distributed data. Panel (a) shows 10,000 samples from a member of the family of distributions used to sample data, in this case with correlation of 0.7 between the two variables. Panel (b) shows the observed power in simulation as a function of effect size, for simulated data of length 100. The Pearson correlation is most powerful at lower effect sizes, while the CI for effects were the power observed is 0.5 or larger. As the sample size is increased while decreasing effect size to keep power stable, the CI gains in power relative to the Pearson correlation, both for regimes of moderate (0.5, Panel (c)) and low power (0.25, Panel (d)).}
\label{fig:supplementalBetaPower}
\end{figure}

\begin{figure}[H]
\centering

\begin{subfigure}{0.7\textwidth}
\centering
\includegraphics[width=\linewidth]{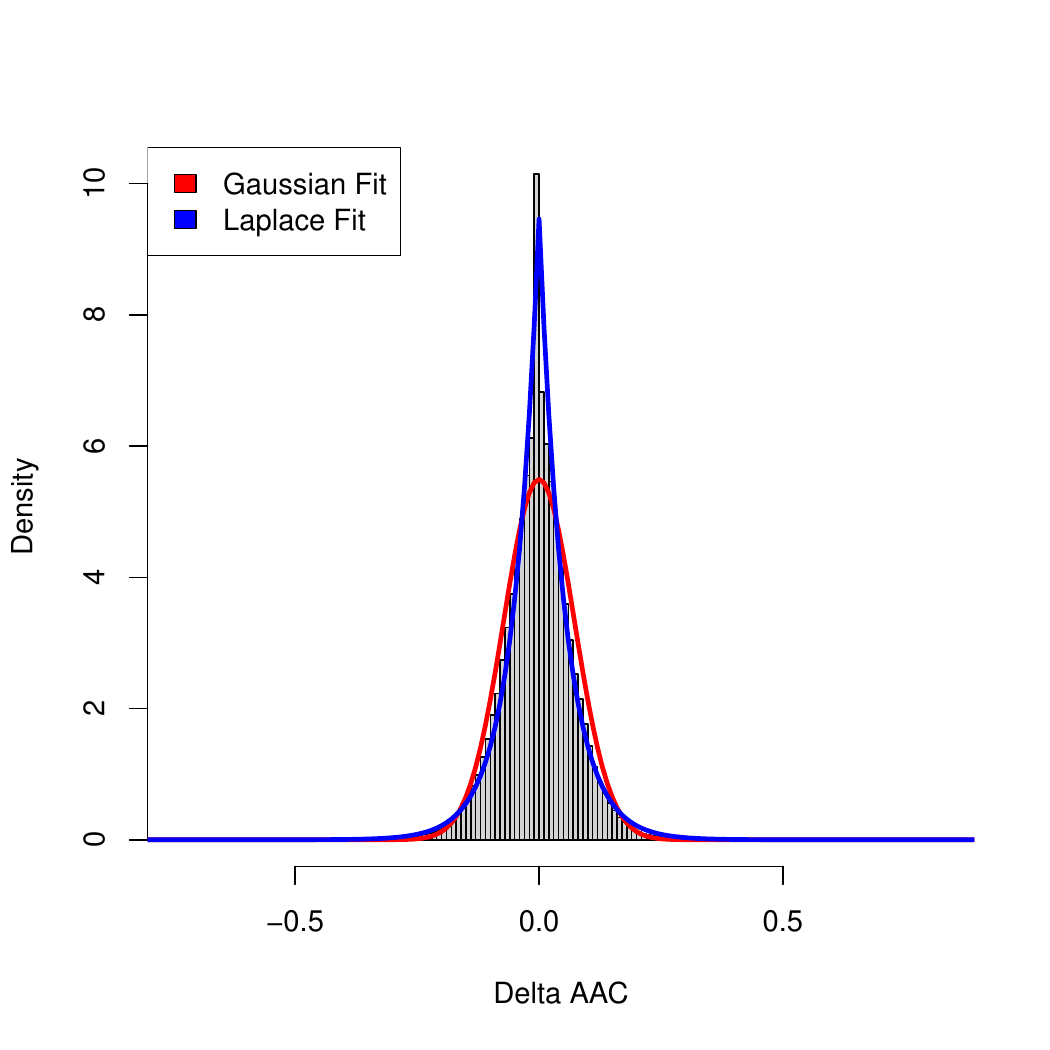}
 \caption{}
 \label{fig:supplementalLaplaceNoisePlot}
\end{subfigure}\\
\begin{subfigure}{0.7\textwidth}
\centering
\includegraphics[width=\linewidth]{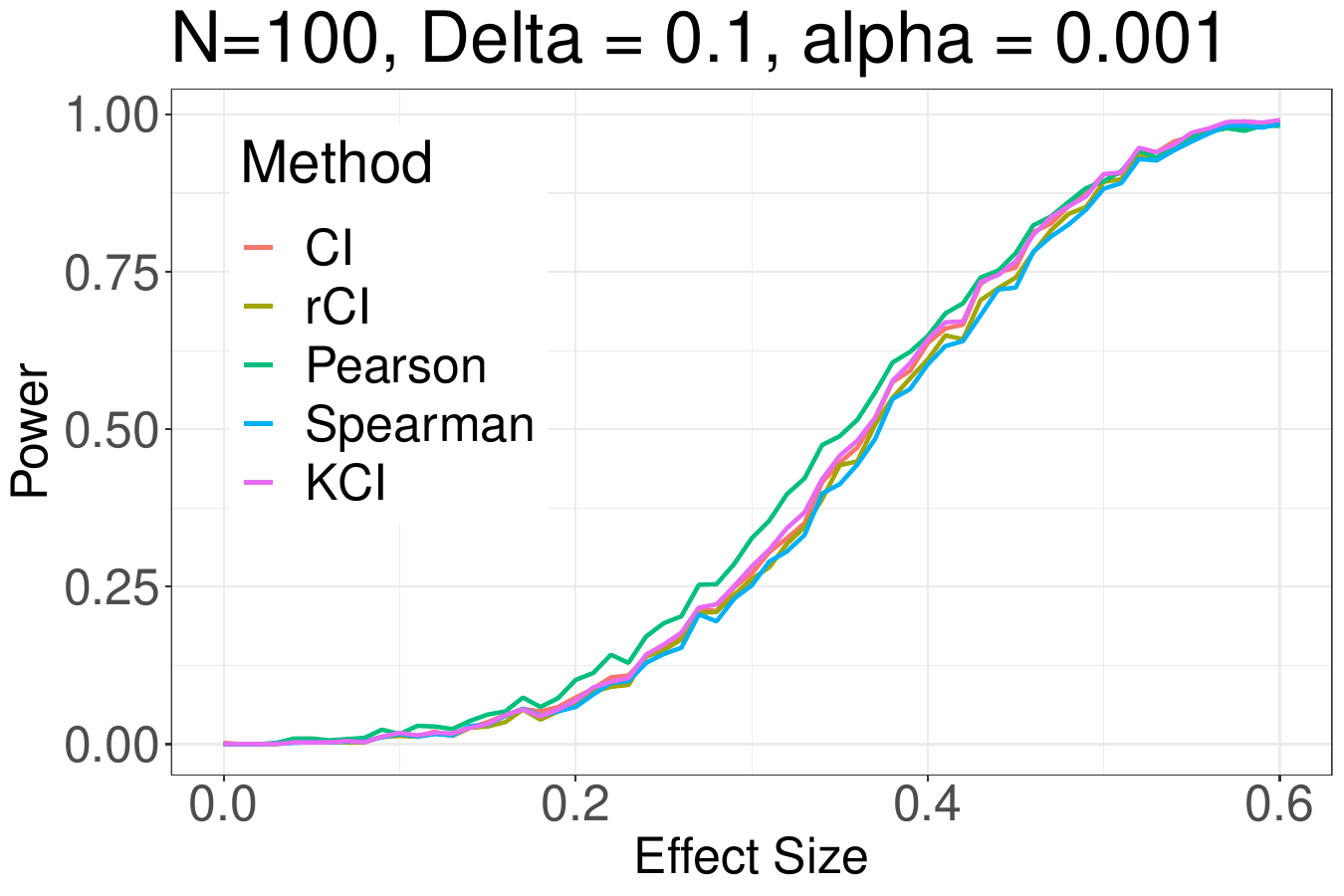}
 \caption{}
 \label{fig:supplementalBetaPowerNoise}
\end{subfigure}\\
\caption{The distribution of differences in AAC measurements between replicates across all the datasets (Delta AAC) (a), with the MLE Gaussian and Laplace distribution plotted over the histogram of observed values. Power simulations for Beta distributed data were repeated with noise sampled from the MLE laplacian to one of the vectors added to the data, for expected correlations prior to adding noise of 0-0.6. While all correlation coefficients displayed lower power than in the noise-free case, surprisingly the Pearson correlation was least affected. }
\end{figure}
\printbibliography


\end{document}